\input harvmac
\noblackbox

\def\RR{${\cal R}\otimes {\cal R}$}

\def\tr{{\rm tr}}
\def\Tr{{\rm Tr}}

\def\Re{\rm Re}


 \font\cmss=cmss10
\font\cmsss=cmss10 at 7pt
\def\rlx{\relax\leavevmode}
\def\inbar{\vrule height1.5ex width.4pt depth0pt}
\def\IC{\relax\,\hbox{$\inbar\kern-.3em{\rm C}$}}
\def\IN{\relax{\rm I\kern-.18em N}}
\def\IP{\relax{\rm I\kern-.18em P}}
\def\ZZ{\rlx\leavevmode\ifmmode\mathchoice{\hbox{\cmss Z\kern-.4em Z}}
 {\hbox{\cmss Z\kern-.4em Z}}{\lower.9pt\hbox{\cmsss Z\kern-.36em Z}}
 {\lower1.2pt\hbox{\cmsss Z\kern-.36em Z}}\else{\cmss Z\kern-.4em
 Z}\fi}
\def\IZ{\relax\ifmmode\mathchoice
{\hbox{\cmss Z\kern-.4em Z}}{\hbox{\cmss Z\kern-.4em Z}}
{\lower.9pt\hbox{\cmsss Z\kern-.4em Z}}
{\lower1.2pt\hbox{\cmsss Z\kern-.4em Z}}\else{\cmss Z\kern-.4em
Z}\fi}

\def\narrowplus{\kern -.04truein + \kern -.03truein}
\def\narrowminus{- \kern -.04truein}
\def\narrowminussub{\kern -.02truein - \kern -.01truein}

\def\kh{K{\"a}hler}

\def\frac#1#2{{#1\over #2}}

\def\R{{\rm R}}

\def\IZ{\relax\ifmmode\mathchoice
{\hbox{\cmss Z\kern-.4em Z}}{\hbox{\cmss Z\kern-.4em Z}}
{\lower.9pt\hbox{\cmsss Z\kern-.4em Z}}
{\lower1.2pt\hbox{\cmsss Z\kern-.4em Z}}\else{\cmss Z\kern-.4em
Z}\fi}
\def\IB{\relax{\rm I\kern-.18em B}}
\def\IC{{\relax\hbox{$\inbar\kern-.3em{\rm C}$}}}
\def\ID{\relax{\rm I\kern-.18em D}}
\def\IE{\relax{\rm I\kern-.18em E}}
\def\IF{\relax{\rm I\kern-.18em F}}
\def\IG{\relax\hbox{$\inbar\kern-.3em{\rm G}$}}
\def\IGa{\relax\hbox{${\rm I}\kern-.18em\Gamma$}}
\def\IH{\relax{\rm I\kern-.18em H}}
\def\II{\relax{\rm I\kern-.18em I}}
\def\IK{\relax{\rm I\kern-.18em K}}
\def\IP{\relax{\rm I\kern-.18em P}}

\font\cmss=cmss10 \font\cmsss=cmss10 at 7pt
\def\IR{\relax{\rm I\kern-.18em R}}

\def\1{{\bf 1}}
\def\3{{\bf 3}}
\def\7{{\bf 7}}
\def\2{{\bf 2}}
\def\8{{\bf 8}}

%

%
%
\def\eqnn#1{\xdef #1{(\secsym\the\meqno)}\writedef{#1\leftbracket#1}%
\global\advance\meqno by1\wrlabeL#1}
\def\eqna#1{\xdef #1##1{\hbox{$(\secsym\the\meqno##1)$}}
\writedef{#1\numbersign1\leftbracket#1{\numbersign1}}%
\global\advance\meqno by1\wrlabeL{#1$\{\}$}}
\def\eqn#1#2{\xdef #1{(\secsym\the\meqno)}\writedef{#1\leftbracket#1}%
\global\advance\meqno by1$$#2\eqno#1\eqlabeL#1$$}

\def\Re{\rm Re}

\def\tr{{\rm tr}}
\def\Tr{{\rm Tr}}

\def\lam16{\lambda^{16}}

\def\lr { \lref}

\parskip=0pt plus 1pt
\parindent 25pt
\tolerance=10000

\lref\gsw{M. Green, J. Schwarz and E. Witten, {\it Superstring
theory}, Cambridge U. Press (1987).}


\lr\joe{J. Polchinski, {\it Dirichlet-Branes and Ramond-Ramond
Charges}, hep-th/9510017; Phys. Rev. Lett. {\bf 75} (1995) 4724.}
\lr\li{M. Li, {\it Boundary States of D-branes and Dy-Strings},
hep-th/9510161;  Nucl. Phys. {\bf B460} (1996) 351.}
\lr\leigh{R.G. Leigh, {\it Dirac-Born-Infeld action from Dirichlet
sigma model},
 Mod. Phys. Lett. {\bf A4} (1989) 2767.}
\lr\bac{C. Bachas, {\it D-brane Dynamics},  hep-th/9511043; Phys. Lett. {\bf B374}
(1996) 37.}

\lr\fradtseyt{E.S.  Fradkin and A.  Tseytlin, {\it Non-linear
Electrodynamics from Quantized Strings}, Phys. Lett. {\bf B163}
(1985) 123.}
\lr\callan{C.G.  Callan,  C. Lovelace,  C.R.  Nappi and S.A. Yost,
{\it Loop Corrections to Superstring Equations of Motion},
 Nucl. Phys. {\bf 308} (1988) 221.}
\lr\douglasa{M.  Douglas, {\it Branes within branes}, hep-th/9512077.}
\lr\bershvaf{M.  Bershadsky, V.  Sadov and C.  Vafa,
 {\it D-branes and topological field theories},  hep-th/9511222;
 Nucl. Phys.  {\bf
 B462} (1996) 420.}
\lr\ghm{M.B.  Green, J.A.  Harvey and G.  Moore,
 {\it I-brane inflow and anomalous couplings on D-branes},
 hep-th/9605033;
  Class. Quant. Grav.  {\bf 14} (1997) 47.}
\lr\cheunga{Y-K. E. Cheung and Z.  Yin, {\it Anomalies, branes and
currents},  hep-th/9710206;  Nucl. Phys. {\bf B517} (1998) 69.}

\lr\alvarezginsparg{L. Alvarez-Gaum{\'e} and P. Ginsparg,
{\it The structure of gauge and gravitational anomalies},
Ann. Phys. {\bf 161} (1985) 423; Erratum-ibid. {\bf 171} (1986) 233.}

\lr\ovrut{ K. Forger, B. A. Ovrut, S.J. Theisen, D. Waldram, {\it
Higher-Derivative Gravity in String Theory}, hep-th/9605145;
 Phys.Lett. {\bf B388} (1996) 512.}


\lr\polch{J. Polchinski, {\it TASI Lectures on D-Branes},
 hep-th/9611050.}

\lr\rev{C.Bachas, {\it Lectures on D-branes}, hep-th/9806199.}


\lr\dasguptaa{K. Dasgupta, D.P. Jatkar and S. Mukhi,
{\it Gravitational couplings and $Z_2$ orientifolds},
hep-th/9707224;
Nucl. Phys. {\bf B523} (1998) 465.}
\lr\dasguptab{K. Dasgupta and S. Mukhi, {\it Anomaly Inflow on
orientifold planes}, hep-th/9709219;
 J. High Energy Phys. {\bf 9803} (1998) 004.}


\lr\vafa{C.  Vafa, {\it Evidence for F-theory},  hep-th/9602022;
 Nucl. Phys.
{\bf B469} (1996) 403.}

\lr\greengaberd{M. Gaberdiel and M. Green, {\it
 An $SL(2,\IZ)$ anomaly in IIB supergravity and its F-theory
 interpretation}, hep-th/9810153; 
J. High Energy Phys. {\bf 9811} (1998) 026. }


\lr\witnorm{
E. Witten, {\it Five-Brane Effective Action In M-Theory},
hep-th/9610234~;
J.Geom.Phys. {\bf 22} (1997) 103-133.
}

\lr\ruben{D. Freed, J. A. Harvey, R. Minasian and  G. Moore,
{\it Gravitational Anomaly Cancellation for M-Theory Fivebranes},
hep-th/9803205~; Adv.Theor.Math.Phys. {\bf 2} (1998) 601-618.}

\lr\bonora{
L. Bonora, C. S. Chu and  M. Rinaldi, {\it Perturbative Anomalies of
the M-5-brane}, hep-th/9710063~; JHEP {\bf 9712} (1997) 007,
{\it Anomalies and Locality in Field Theories and M theory}, hep-th/9712205.
}


\lr\dfkz{J. P. Derendinger, S. Ferrara, C. Kounnas and F. Zwirner, {\it On
loop corrections to string effective field theories: field-dependent
gauge couplings and $\sigma$-model anomalies}, Nucl. Phys. {\bf B372}
(1992) 145. }
\lref\lco{G. L. Cardoso and B. A. Ovrut, {\it A Green-Schwarz mechanism
for $D=4$, ${\cal N}=1$ supergravity anomalies}, Nucl. Phys. {\bf
B369} (1992) 351;
{\it Coordinate and \kh sigma model anomalies and their cancellation
in string effective field theories},
hep-th/9205009;  Nucl. Phys. {\bf B392} (1993)
315.}
\lr\carlustov{G.L. Cardoso, D. Lust, and
B.A. Ovrut, {\it Moduli dependent non-holomorphic
contributions of massive states to gravitational
couplings and $C^2$ terms in $Z(N)$ orbifold compactification},
hep-th/9410056;
Nucl. Phys. {\bf B436} (1995) 65.}
\lref\kaplouis{V. Kaplunovsky and J. Louis, {\it On gauge couplings in string theory},
 hep-th/9502077;
Nucl. Phys. {\bf B444} (1995) 191. }
\lref\kaplouisp{V. Kaplunovsky and J. Louis, {\it Field dependent
gauge couplings in locally supersymmetric
effective quantum field theories}, hep-th/9402005; Nucl. Phys. {\bf B422}
(1994) 57.}


\lr\morales{J.F.  Morales, C.A. Scrucca and M. Serone,
 {\it Anomalous couplings for D-branes and O-planes}, hep-th/9812071.}
\lr\craps{B.  Craps and F.  Roose, {\it Anomalous
 D-brane and orientifold couplings from the boundary state},
 hep-th/9808074; Phys.Lett. {\bf B445} (1998) 150-159; {\it
 (Non-)Anomalous D-brane and O-plane couplings: the 
normal bundle}, hep-th/9812149.}
\lr\stefanski{B.  Stefanski, {\it Gravitational Couplings of D-branes
and O-planes}, hep-th/9812088.}

\lr\garmyers{M.R. Garousi and R.C. Myers,
{\it Superstring scattering from D-branes},  hep-th/9603194;
Nucl. Phys. {\bf B475} (1996) 193-224;
{\it World-volume interactions on D-branes}, hep-th/9809100; Nucl.Phys. {\bf B542} (1999) 73-88.}
\lr\hashkleb{A. Hashimoto, I. R. Klebanov, {\it Decay of Excited
D-branes},   hep-th/9604065; Phys. Lett. {\bf B381} (1996) 437-445; {\it
Scattering of Strings from D-branes},   hep-th/9611214;
Nucl. Phys. Proc. Suppl. {\bf 55B} (1997) 118-133. }

\lr\kirb{
C. Bachas and  E. Kiritsis, {\it $F^4$  Terms in ${\cal N}=4$ String Vacua},
 hep-th/9611205; 
 Nucl.Phys.Proc.Suppl. {\bf 55B} (1997) 194-199. 
}

\lr\ab{I. Antoniadis and C. Bachas, {\it Branes and the Gauge
Hierarchy}, hep-th/9812093; Phys. Lett. {\bf B450} (1999) 83. }


\lr\insta{
H. Ooguri and C. Vafa, {\it Summing up D-Instantons},
hep-th/9608079;
Phys. Rev. Lett. {\bf 77} (1996) 3296-3298.
}
\lr\inst{
C. Bachas, C. Fabre, E. Kiritsis, N. A. Obers and  P. Vanhove,
{\it Heterotic / type I duality and D-brane instantons},
hep-th/9707126;
Nucl. Phys. {\bf B509} (1998) 33-52.
}
\lr\instn{
E. Kiritsis and  N. Obers, {\it Heterotic/Type-I Duality in D $<10$
 Dimensions,
 Threshold Corrections and D-Instantons},   hep-th/9709058;
 J. High Energy Phys. {\bf 9710} (1997) 004.
}
\lr\instt{C. Bachas, {\it Heterotic versus Type I},
hep-th/9710102;
Nucl. Phys. Proc. Suppl. {\bf 68}  (1998) 348-354.}
\lr\instb{
I. Antoniadis, B. Pioline and  T. R. Taylor, {\it Calculable
$e^{-1/\lambda}$  Effects},   hep-th/9707222; 
Nucl. Phys. {\bf B512}  (1998) 61-78.
}

\lr\lerche{W. Lerche and S. Stieberger, {\it Prepotential, Mirror Map and
F-Theory on K3},  hep-th/9804176; Adv. Theor. Math. Phys.  {\bf 2} (1998)
1105.}

\lr\kristin{K. Foerger and S. Stieberger, {\it Higher Derivative Couplings
and Heterotic-Type I Duality in Eight Dimensions}, hep-th/9901020. }

\lr\lerch{W. Lerche, S. Stieberger and  N. P. Warner, {\it
Prepotentials from Symmetric Products}, hep-th/9901162.}

\lr\lerc{W. Lerche, S. Stieberger and  N. P. Warner, {\it
Quartic Gauge Couplings from K3 Geometry}, hep-th/9811228. }

\lr\harvmoor{J. Harvey and G. Moore,
{\it Fivebrane Instantons and $R^2$ couplings in ${\cal N}=4$ String
 Theory},   hep-th/9610237;
 Phys. Rev. {\bf D57} (1998) 2323-2328.}
\lr\gkkopp{A. Gregori, E. Kiritsis, C. Kounnas, N. A. Obers,
P. M. Petropoulos and B. Pioline,
{\it $R^2$ Corrections and Non-perturbative Dualities of
${\cal N}=4$ String ground states},
hep-th/9708062; Nucl. Phys. {\bf B510} (1998) 423-476.}

\lr\ggv{M. B. Green, M. Gutperle and P. Vanhove,
{\it One loop in eleven dimensions},  hep-th/9706175;  Phys. Lett. {\bf B409} (1997)
177-184.}
\lr\ggk{M.B. Green,  M. Gutperle and H. Kwon, {\it Sixteen Fermion
 and Related
Terms in M-theory on $T^2$}, hep-th/9710151;  Phys. Lett. {\bf B421} (1998) 149.}

\lr\tseytlina{A.  Tseytlin, {\it Self-duality of Born-Infeld action
and Dirichlet 3-brane of type IIB superstring theory},
hep-th/9602064;
Nucl. Phys. {\bf B469} (1996) 51-67.}
\lr\greengutd{M.B. Green and M.  Gutperle, {\it Comments on D3-branes},
 hep-th/9602077; Phys. Lett. {\bf B377} (1996) 28-35.}
\lr\gibbons{G.W.  Gibbons and Rasheed, {\it SL(2,R) Invariance of
Non-Linear Electrodynamics Coupled to An Axion and a Dilaton},
hep-th/9509141;
Phys. Lett. {\bf B365} (1996) 46-50.}

\lr\deroo{M. de Roo,  {\it Matter coupling in ${\cal N}=4$ Supergravity},
Nucl. Phys. {\bf B255} (1985) 515.}
\lr\bergshoeff{ E.  Bergshoeff, I. Koh and E. Sezgin,  {\it Coupling
of Yang Mills to ${\cal N}=4$, $d=4$ Supergravity}, Phys. Lett. {\bf 155B}
(1985) 71.}

\lr\theisen{S. Theisen, {\it Fourth order supergravity},
Nucl. Phys. {\bf B263} (1986) 687; Addendum-ibid. {\bf B269} (1986) 744.}

\lr\greenguta{M.B. Green and M. Gutperle, {\it D-particle bound states
and the D-instanton measure},  hep-th/9711107; J. High Energy Phys. {\bf 9801} (1998) 005.}
\lr\greenguttwo{M.B. Green and M. Gutperle, {\it D-instanton partition
functions},  hep-th/9804123;
 Phys. Rev. {\bf D58} (1998) 046007.}
\lr\yi{P. Yi, {\it Witten Index and Threshold Bound States of
 D-Branes},   hep-th/9704098; Nucl. Phys.
{\bf B505} (1997) 307-318.}
\lr\sethi{S. Sethi and M. Stern, {\it D-Brane Bound States Redux},
 hep-th/9705046; Commun. Math. Phys.
 {\bf 194} (1998) 675-705.}

\lr\ramwal{S. Ramgoolam and D. Waldram, {\it Zero-Branes on a Compact
Orbifold},   hep-th/9805191; J.High Energy Phys. {\bf 9807} (1998) 009.}
\lr\douglas{M. Douglas, private communication.}


\lr\eisen{L.P. Eisenhart,
{\it Riemannian geometry}, Princeton U. Press,  1926.  }
\lr\koba{S. Kobayashi and K. Nomizu, {\it Foundations of
Differential Geometry}, J. Wiley \& Sons, New York 1969, vol.2.}


\lr\witvafa{C. Vafa and E. Witten, {\it A Strong Coupling Test of
S-Duality},  hep-th/9408074; Nucl. Phys.
 {\bf B431} (1994) 3-77.}
\lr\wittverl{E. Witten, {\it On S-duality in Abelian gauge Theory},
  hep-th/9505186; Mod. Phys. Lett. {\bf A10} (1995) 2153;
E. Verlinde, {\it Global Aspects of Electric-Magnetic Duality},
hep-th/9506011; Nucl. Phys. {\bf B455} (1995) 211.}

\lr\horavwit{P. Horava and E. Witten, {\it Heterotic and Type I String
  Dynamics from Eleven Dimensions}, hep-th/9510209; Nucl. Phys. {\bf
  B460} (1996) 506; {\it Eleven-Dimensional
Supergravity on a Manifold with Boundary}, hep-th/9603142;
Nucl. Phys. {\bf B475} (1996) 94.
}


\lr\hulltown{C.M. Hull and P. K. Townsend, {\it Unity of Superstring
Dualities }, hep-th/9410167; Nucl. Phys. {\bf B438} (1995) 109-137}
\lr\duffkhuri{M.J. Duff and R. R. Khuri, {\it Four-Dimensional
String/String Duality},
hep-th/9305142; Nucl. Phys. {\bf B411} (1994) 473}
\lr\duff{M.J. Duff,
{\it Strong/Weak Coupling Duality from the Dual String},
 hep-th/9501030; Nucl. Phys. {\bf B442} (1995) 47-63.}
\lr\schwarz{J.H. Schwarz,  {\it An SL(2,Z) Multiplet of
Type IIB Superstrings},  hep-th/9508143; Phys. Lett. {\bf B360} (1995) 13-18;
Erratum-ibid. {\bf B364} (1995) 252.}
\lr\schwarza{J.H. Schwarz, {\it The Power of M-Theory},
hep-th/9510086; Phys. Lett.  {\bf B367} (1996) 97-103.}
\lr\aspinwall{P.S. Aspinwall, {\it Some relationships between
dualities in string theory},  hep-th/9508154
  Nucl. Phys. Proc. Suppl. {\bf 46} (1996) 30-38.}

\lr\lerche{ W. Lerche and  S. Stieberger, {\it
Prepotential, Mirror Map and F-Theory on K3}, hep-th/9804176.}


\Title{\vbox{\baselineskip12pt
\hbox{hep-th/9903210;  LPTENS-99-07; DAMTP-1999-14}
}}
{\vbox{
\centerline{Curvature terms in D-brane actions}
\smallskip
\centerline{and their M-theory origin}}}
\medskip
 \centerline{Constantin P.  Bachas\foot{email:
bachas@physique.ens.fr},  Pascal Bain\foot{email:
bain@physique.ens.fr}}
\vskip 0.05in
\centerline{\it Laboratoire de Physique Th{\'e}orique de l'
{\'E}cole Normale Sup{\'e}rieure
\foot{Unit{\'e}
mixte  du CNRS,  UMR8549 .}}
\centerline{\it 24 Rue Lhomond, 75231 Paris Cedex, France}
\smallskip
\centerline{and}
\smallskip
\centerline{Michael B.  Green\foot{email: M.B.Green@damtp.cam.ac.uk}}
\vskip 0.05in
\centerline{\it Department of Applied Mathematics and
Theoretical Physics}
\centerline{\it Silver Street,  Cambridge CB3 9EW, UK}

{\baselineskip 12pt plus 2pt minus 2pt
{\vskip 0.4 cm
We derive the  complete $($curvature$)^2$  terms of  effective D-brane actions,
for arbitrary  ambient geometries and world-volume embeddings,
at lowest order (disk-level) in  the string-loop expansion.
These terms reproduce  the $o(\alpha^{\prime\;2})$ corrections to
string scattering amplitudes, and are consistent with duality conjectures.
In the particular case of the D3-brane with trivial normal bundle,
 considerations of $SL(2,\IZ)$
invariance lead to a complete sum of D-instanton corrections for  both
the parity-conserving and the parity-violating parts of the effective
action. These corrections are required for the cancellation of the modular
anomalies of massless modes, and are consistent with the
absence of chiral anomalies in the intersection domain of pairs of
D-branes. We also show
that the  parity-conserving part of the
 non-perturbative $R^2$ action follows from a one-loop
quantum calculation in the six-dimensional world-volume of the M5-brane
compactified on a two-torus.}}

\medskip
\Date{2/99}

\newsec{Introduction}

The dynamics of $p$-branes is well-approximated by effective
world-volume $(p+1)$-dimensional field theories of various kinds when
$p<5$.  In the case of Dp-branes the effective  world-volume actions consist of
the sum of  a Wess--Zumino (WZ) part  that is parity-violating, and of
a parity-conserving part whose first term is the
 Dirac--Born--Infeld (DBI) action (see
\refs{\polch,\rev}  for reviews).

The WZ term describes the coupling of a Dp-brane to the
bulk Ramond--Ramond (\RR) fields and can be motivated in a variety of
manners \refs{\joe,\callan,\li,\douglasa,\bershvaf,\ghm,\cheunga} .
It has the structure of an anomaly cancelling term  and can indeed
be determined by requiring the consistent cancellation of the chiral
gauge and gravitational anomalies that arise when pairs of D-branes
intersect in certain configurations \refs{\ghm,\cheunga}.
Such terms have a simple topological description which results in
the expression for the $(p+1)$-form WZ term
\eqn\wzdef{\int_{M^{(p+1)}} {\cal L}^{(p) {disk} }_{WZ}  =
T_{(p)}\int_{M^{(p+1)}}
  C \wedge \tr_N\left(e^{2\pi\alpha^\prime  F}\right)
 \wedge \left(\hat {\cal A}(4\pi^2\alpha^\prime R_T) \over \hat {\cal
  A}(4\pi^2\alpha^\prime
R_N)\right)^{1/2},}
where $T_{(p)}$ is the tension of the Dp-brane and
the trace is in the fundamental representation of the
 world-volume gauge group, $U(N)$.
The expression $\hat{\cal  A}$ is the
  Dirac `roof'  genus
and its square root has the expansion in even powers of the curvature two-form,
\eqn\sqdir{\sqrt{\hat {\cal A}(R)} = 1  - {1\over 48} p_1(R) + {1\over 2560}
  p_1^2 (R) -
 {1\over 2880} p_2(R) + \dots,}
where $p_1$ and $p_2$ are the first two Pontryagin classes.
 The components of the curvature with
tangent-space indices  are denoted by $R_T$  in \wzdef\ while  $R_N$ denotes the
components in the normal bundle (the definitions of our notation are
 made precise in the appendices).
   The \RR\ $(p+1)$-form  potentials pulled back to the Dp-brane
are contained in $C = C^{(0)} + C^{(1)} \dots + C^{(9)}$, where the
odd forms contribute in the IIA theory and the even ones in the IIB
theory,
 The structure of \wzdef\   may also
be determined by considering the scattering of gravitons on a
world-sheet with disk topology \refs{\craps,\stefanski,\morales},
 which we will refer to as the `tree-level' contribution.  As usual,
since these  terms are responsible for the cancellation of anomalies 
they receive no perturbative corrections.

The DBI action for a single Dp-brane \refs{\fradtseyt,\leigh}
encapsulates the exact dependence
 on the field strength of the Born--Infeld vector potential for
 constant field strengths, and can be fixed by considerations of
T-duality and Lorentz covariance \refs{\bac}
\eqn\bi{
\int_{M^{(p+1)}} {\cal L}_{DBI}^{(p)} = T_{(p)}
\int_{M^{(p+1)}}  \;e^{-\phi}  \sqrt {{\rm det}
 \,\Bigl( (G_{\mu \nu}+B_{\mu \nu}) \partial_\alpha
 Y^\mu \partial_\beta Y^\nu + 2\pi\alpha^\prime F_{\alpha\beta} \Bigr)} .
}
Here and in the
following, we will use Greek indices from the middle of the alphabet
$(\mu , \nu,...)$ for the space-time  coordinates, Greek letters from
the beginning of the alphabet $(\alpha, \beta ...)$ for the
 world-volume coordinates, and Latin letters $(a,b,...)$ for the normal
 bundle (see  appendix A for more details). In contrast to the WZ
 action,  the expression \bi\ does not include  gravitational-curvature
 effects induced by the ambient geometry and/or by non-trivial
 world-volume embeddings.
Such effects will generically require
 the addition of terms that again involve powers of the Riemann
 curvature with nontrivial values in both the tangent bundle and the
 normal bundle. In section 2  we will determine
 the structure of these  terms by considering the scattering of
 gravitons and of ripples (open strings)
in the world-volume of the D-brane. A residual ambiguity will be fixed
 by invoking  heterotic/type II duality in
 six dimensions.  Our analysis in section 2
will be restricted to the lowest-order terms,
 which arise from world-sheets with the topology of a disk  and which
 are quadratic in the  curvature expansion.   For
 simplicity we will also assume constant
 dilaton and  \RR\  backgrounds, as well as vanishing  $B_{\mu \nu}$ and
 $F_{\alpha\beta}$ fields, but we allow for arbitrary ambient geometries and
 world-volume embeddings.

In section 3 we will consider the particular case of D3-branes.
 We will explain why
 the curvature-squared terms receive  non-perturbative D-instanton
 corrections  that are necessary in order to ensure that the
 type IIB theory maintains the  requisite $SL(2,\IZ)$ invariance
 \refs{\hulltown, \schwarz} in the presence of D3-branes.
 The terms at issue are of the form
 $f(\tau,\bar \tau) R^2$, where $R$ denotes the components of the
 Riemann curvature and the  function $f(\tau,\bar \tau)$ has specified
 properties under modular transformations
  of the complex scalar field of the IIB theory,
\eqn\scalar{
\tau \equiv  \tau_1 + i\tau_2 = {1\over \lambda_s^B}
\left[ C^{(0)} + i e^{-\phi}\right],
}
 where $C^{(0)}$ is the \RR\ scalar and $\phi$ is the type IIB
 dilaton.\foot{The D-brane tensions and the gravitational
 coupling are given by
$2\kappa^2_{(10)} = \lambda_s^2\alpha^{\prime\ 4} (2\pi)^7$ and
$\kappa_{(10)}^2 T_{(p)}^2 =
 \pi (4\pi^2\alpha^\prime)^{{3-p}}$. The symbols $\lambda_s^A$ and
 $\lambda_s^B$ will be used to distinguish the string couplings in the
 IIA and IIB theories, respectively, and the 
 dilaton $\phi$ has zero vacuum expectation
 value.}
In the special case of a trivial normal bundle we will see that the 
complete expression for the $R^2$ terms in the
 D3-brane action will be given by,
\eqn\fullconj{\int_{M^{(4)}} {\cal L}_{R^2}^{(3)} =
 {1\over 16\pi^2}
\, \int_{M^{(4)}} {\Re}\Bigl[   \log \eta(\tau)\; {\rm tr} (\R\,{\wedge^*\R} -
i \R\wedge \R)\Bigr]  ,}
where $\eta(\tau)$ is the Dedekind function, $\R \equiv {1 \over 2}
R_{\alpha\beta}\;  d\zeta^\alpha \wedge d\zeta^\beta$ is the
$SO(1,3)$-valued curvature two-form, and  $^*\R \equiv {1\over
4}\sqrt{g}\; \epsilon_{\alpha\beta}^{\ \ \ \gamma\delta}\;
R_{\gamma\delta}\; d\zeta^\alpha \wedge d\zeta^\beta$ is
its Hodge dual.
The expression \fullconj\ has a small-coupling expansion ($\tau_2 \to
\infty$)
 that contains   the
`tree-level' terms that arise from the disk calculations in section 2
together
 with an infinite series of D-instanton corrections.

One argument for the form \fullconj\ of the D3-brane action,  based
on type IIA/F-theory duality \refs{\dasguptaa}, will be described in
the first part of section 3. The F-theory background contains  24
$(p,q)$ seven-branes, each making an equal contribution to the $R^2$ 
terms of the
effective low-energy action. This universal piece of the seven-brane
action is  related by T-duality to the $R^2$ terms of the D3-brane action.
 On the type-IIA side, on the other hand,
these same  terms come from a  one-loop amplitude which was
 computed in  \refs{\harvmoor,\gkkopp}. Comparing the two sides leads to
expression \fullconj\ .

A different argument for the structure of \fullconj\ follows by
careful consideration of its properties under $SL(2,\IZ)$
transformations.
We will see in the second part of section 3 that the conjectured
action  transforms
under $SL(2,\IZ)$ transformations in precisely the appropriate manner
to cancel the $\tau$-dependent anomalous transformation of the
determinant of the massless world-volume fields.  This anomaly
 results from the chiral couplings of the
world-volume fields to the $SL(2,\IR)$ Noether current.  A
$\tau$-independent modular anomaly remains, but it  is the integral of a
topological density  which  integrates to an integer  multiple of
$2\pi$ in consistent backgrounds

The emphasis in section 4 is on the relationships  between the
nonperturbative
curvature terms in the D3-brane of the IIB theory and properties of
the M5-brane compactified on $T^2$.  Besides specializing  to the
case of a trivial normal bundle, we also  only consider the parity
conserving terms.
We will make use of the
various dualities that relate M-theory on $T^2$ to the type IIA and
IIB theories compactified on $S^1$ \refs{\schwarza,\aspinwall}
to relate the D3-brane  to higher-dimensional
branes.  T-duality on the  circle
relates the IIB theory to the IIA theory,  and  converts an
unwrapped D3-brane into a wrapped D4-brane.  The instanton terms in
the D3-brane are now interpreted as the world-lines of  D-particles
which form bound states with  the D4-brane
\refs{\polch}. These are the expected threshold bound states
of a quantum mechanical system with ${\cal N}=(4,4)$ supersymmetry.
As we will  see, the exact expression for the D-instanton measure that is
 extracted from the function $f(\tau,\bar \tau)$ is consistent with
the fact that the Witten index for this D-particle--D4-brane is one.
The D4-brane in turn may be interpreted as the M5-brane wrapped
around the circular eleventh dimension of M-theory. The D-particle bound
states are then simply the Kaluza--Klein excitations in the double
dimensional reduction of the M5-brane.  In this picture the $R^2$
terms can be deduced by a calculation of a one-loop effect in
the world-volume of the compactified M5-brane which will be described
in section 4. This is
 analogous to the manner in which the $R^4$ terms
of the bulk theory were  deduced in \ggv .

Section 5 summarizes our conclusions, and outlines  how our arguments
could be extended in several directions. We discuss in particular why
non-perturbative $R^4$ terms are expected in the world-volume action of a
seven-brane, and  how the analysis of sections 3 and 4
could  be extended to account for  non-trivial normal bundles. Finally we
comment on  the relevance of our results in the context of
 the AdS/CFT correspondence.

\newsec{Disk-level curvature terms}

In this section we will extract the $R^2$ terms of the classical D-brane
action from  disk-level  scattering amplitudes, combined with a duality
argument and the requirement  of reparametrization invariance.
We begin by reviewing the calculation \refs{\garmyers,\hashkleb} of
the
 disk diagram for (a) the
elastic scattering of a  graviton off a  D-brane, (b) the absorption of a
graviton whose energy is carried away by two ripples (represented by
open strings)  in the
world-volume of the D-brane,
and (c) the elastic scattering of two such world-volume ripples.
These processes can all be conveniently expressed in terms of the standard
amplitude for the scattering of four open-string gauge bosons
 \refs{\gsw}
\eqn\opens{A(\zeta_1,k_1;\zeta_2,k_2;\zeta_3,k_3;\zeta_4,k_4)=
 {\cal N}\; K(1,2,3,4)\;
\frac{\Gamma(2\alpha^\prime k_1\!\cdot \!k_2)
\Gamma(2\alpha^\prime k_1\!\cdot \!k_4)}
{\Gamma(1+2\alpha^\prime k_1\!\cdot \!k_2
+ 2\alpha^\prime k_1\!\cdot \!k_4)}
, }
where
$k_r$ and $\zeta_r$ are momentum  and polarization vectors,  $\sum
k_r = 0$ , and ${\cal N}$ is a normalization constant.
 The kinematic factor is defined in the usual manner
\eqn\knidef{
K(1,2,3,4) =-16\; t_8^{\mu _1 \mu _2 \mu _3 \mu _4
\mu _5 \mu _6 \mu _7 \mu _8} k_{1\mu _1} \zeta_{1\mu _2}k_{2\mu _3}
\zeta_{2\mu _4}k_{3\mu _5}
 \zeta_{3\mu _6}k_{4\mu _7} \zeta_{4\mu _8}}
with  $t_8$ the well-known  eighth-rank tensor, which is antisymmetric
in each of the four pairs of consecutive indices  and symmetric under
interchange of two pairs.

Equation \opens\  can be interpreted as the amplitude for
  any of the processes (a), (b) or (c) that  describe the
  scattering of closed or open strings in the background of a
  static Dp-brane oriented along the first $p$ spatial
  directions.
The massless brane excitations correspond to
external open strings with momentum $k_r$ restricted to the tangent hyperplane.
Normal and tangent open-string polarizations describe, respectively, the
transverse oscillations of the D-brane, and the gauge bosons living on
its world-volume. Amplitudes with an external  graviton
of momentum $p_\mu $ and polarization  $\varepsilon_{\mu \nu}$  can
be obtained by the formal replacements  \refs{\garmyers,\hashkleb}
\eqn\identdef{
 2k_{1\mu }\rightarrow p_\mu \; , \   2k_{2\mu }\rightarrow
(D\!\cdot \!p)_\mu \; ,  \  \qquad {\rm and} \qquad
 \zeta_{1\mu }\zeta_{2\nu} \rightarrow
\varepsilon_{\mu \lambda}D^\lambda_{\;\nu} \; ,
}
where  the
diagonal matrix $D$ has  a
$+1$ entry for a direction tangent to the world-volume of the static
Dp-brane and $-1$ for a
normal direction. Note that  the constraint $\sum
k_r = 0$ is still obeyed, even though the transverse  momentum 
of the graviton is not
necessarily  conserved in the process. It can be checked that with the
above identifications the  low-energy limit  of
these  amplitudes is  compatible with  the
  Born-Infeld action \bi\  coupled to the type-II supergravity lagrangian.
The comparison fixes the normalization constant
\eqn\norm{
{\cal N} \to  -{1\over 8} T_{(p)}\;\alpha^{\prime 2}\; \prod_{\rm open}
(2\pi\alpha^\prime)^{\mp 1/2}
}
where the product runs over open-string  external legs, and the
sign   depends on
whether the external leg is   a world-volume gauge-boson or a transverse brane
oscillation. Note that our normalization of the  graviton vertex operator,
 $G_{\mu \nu} = \eta_{\mu \nu}+2 \epsilon_{\mu \nu}e^{-ipX}$, differs
from the one in  reference \refs{\garmyers}.

We first consider the elastic scattering  of a graviton,
and expand the corresponding amplitude to subleading order in the  momenta
\eqn\twopt{\eqalign{
A(\epsilon_1,p_1;\epsilon_2,p_2) &= - \frac{1}{8}T_{(p)}
\;\alpha^{\prime 2}\; K(1,2)\;
\frac{\Gamma{(-\alpha^\prime t/4)}\Gamma{(\alpha^\prime q^2)}}
{\Gamma{(1-\alpha^\prime t/4+\alpha^\prime q^2)}} \cr
=&  \frac{1}{2}T_{(p)}\;K(1,2)\;
\left(\frac{1}{q^2 t} + \frac{\pi^2\alpha^{\prime 2}}{24}
+ o(\alpha^{\prime 4})\right)}}
where $q^2=p_1\!\cdot \!D\!\cdot \!p_1/2$ is the square momentum  
flowing along  the
world-volume of the   Dp-brane, and $t=-2p_1\!\cdot \!p_2$ is
the momentum transfer in the transverse directions.
The kinematic factor following from the identifications \identdef\ reads
\eqn\kine{K(1,2)=\left(2q^2\,a_1+\frac{t}{2}\,a_2\right)}
with
\eqn\noname{\eqalign{
a_1&={\rm Tr}(\epsilon_1 \!\cdot \!  D)\,p_1 \!\cdot \!  \epsilon_2  \!\cdot \!
p_1 -p_1 \!\cdot \! \epsilon_2 \!\cdot \!
D \!\cdot \! \epsilon_1 \!\cdot \!  p_2 - p_1 \!\cdot \! \epsilon_2 \!\cdot \!
\epsilon_1  \!\cdot \!  D \!\cdot \!  p_1  \cr
&\ -p_1 \!\cdot \! \epsilon_2  \!\cdot \!  \epsilon_1  \!\cdot \!  D  
\!\cdot \!  p_1  -
p_1 \!\cdot \! \epsilon_2 \!\cdot \! \epsilon_1  \!\cdot \!  p_2 + q^2\,{\rm
Tr}(\epsilon_1 \!\cdot \! \epsilon_2)
+ \Big\{1\longleftrightarrow 2\Big\}}}
\eqn\noname{\eqalign{
a_2&={\rm Tr}(\epsilon_1 \!\cdot \!  D)\,(p_1 \!\cdot \! \epsilon_2 \!\cdot 
\!  D
\!\cdot \!  p_2 + p_2 \!\cdot \!
D \!\cdot \! \epsilon_2 \!\cdot \!  p_1 +p_2 \!\cdot \!  D \!\cdot \! 
\epsilon_2 \!\cdot \!
D \!\cdot \!  p_2)
\cr
&+p_1 \!\cdot \!  D \!\cdot \! \epsilon_1 \!\cdot \!  D \!\cdot \! 
\epsilon_2 \!\cdot \!  D
\!\cdot \!  p_2 -p_2 \!\cdot \!
D \!\cdot \! \epsilon_2 \!\cdot \! \epsilon_1 \!\cdot \!  D \!\cdot \!  p_1
+q^2\,{\rm Tr}(\epsilon_1 \!\cdot \!  D \!\cdot \!  \epsilon_2 \!\cdot \!  D)
\cr
&-q^2\,{\rm Tr}(\epsilon_1 \!\cdot \! \epsilon_2)
-{\rm Tr}(\epsilon_1 \!\cdot \!  D) {\rm Tr}(\epsilon_2 \!\cdot \!
D)\,(q^2-t/4)
+\Big\{1\longleftrightarrow 2 \Big\}}}
Note  that 
a two-point function at tree level usually  vanishes  when the on-shell relations and the
momentum conservation are used. In our case, 
since the momentum is conserved only in the directions parallel
 to the world-volume of the brane, the disk-level two-point function
 does not vanish.

The leading term in the  momentum expansion of the amplitude \twopt\  contains poles
from the exchange of open ($1/q^2$) and closed ($1/t$) string states, 
plus contact
terms,  all consistent with the Dirac-Born-Infeld lagrangian  and the
two-derivative supergravity action.
The subleading term  in the amplitude, on the other hand,    can be attributed
entirely to the ${\cal O}(\alpha^{\prime 2})$ contact interactions of two
gravitons on the world-volume of  the D-brane. Indeed, 
 the trilinear vertices in the bulk as well as the
graviton/ripple mixing on the world-volume are protected by
supersymmetry and should not   receive any string  corrections.
Thus there are no 1PI contributions to  the subleading term
of  the disk amplitude. Note that corrections to the above vertices
 can be of course generated  by field
redefinitions, so that 
 a particular choice of fields is  implicitly assumed  in our
discussion.  For  bulk fields this is the standard choice  in
  which   the higher-derivative corrections
to the  type II  supergravity lagrangian are at least 
$({\rm curvature})^4$, and are of  subleading order compared
to the terms considered here.

   Possible  ${\cal O}(\alpha^{\prime 2})$ corrections to the DBI
action, consistent with  space-time and
world-volume reparametrization invariance, are enumerated  in appendix
B. They depend on various pull-backs of the ambient curvature and on
the second fundamental form of the world-volume, $\Omega$.
At the linearized level around flat space and a static D-brane, 
 $\Omega$  contains an
open-string excitation, and the bulk
Ricci tensor $R_{\mu\nu}= R_{\ \mu \rho \nu}^{\rho}$ is zero. 
As a result (see appendix B)
  terms involving $\Omega$ and $R_{\mu\nu}$ do not contribute to
the two-graviton amplitude \twopt. 
One further ambiguity is  related to the 
 Gauss-Bonnet combination
\eqn\gauss{ {\cal L}_{GB} =  {\sqrt{g}\over 32 \pi^2}  \left(  R_{\alpha \beta
\gamma \delta} {R}^{\alpha \beta \gamma  \delta} - 4 {{\hat R}}_{\alpha
\beta} {{\hat R}}^{\alpha \beta} + {{\hat R}}^2\right) .  }
Here and in the following ${\hat R}_{\alpha\beta}$, ${\hat R}_{ab}$
and $\hat R$ are obtained by contracting tangent indices only.
The Gauss-Bonnet combination
 is a topological invariant in four dimensions, but it is
a total derivative at quadratic  order in all  dimensions.
Hence it  does not contribute to the elastic  scattering of a graviton
off the D-brane. Modulo  all these ambiguities, there exist nine
remaining independent
terms  whose coefficients can be fixed by comparing with the
string-theory amplitude \twopt.

 This straightforward but tedious comparison leads to  the
 following result for the $($curvature$)^2$ corrections to the DBI action, 
\eqn\treelevel{\eqalign{
{\cal L}^{(p)}_{CP-even}  &=  c\; e^{-\phi} {\cal L}_{GB} \; + \;
 T_{(p)} e^{-\phi}  {\sqrt{ g}}\;  \Bigl[    1  \; -\;
 \frac{1}{24}\frac{(4\pi^2\alpha^\prime)^2}{32 \pi^2}\times 
 \cr
& \times  {( R_{\alpha
\beta \gamma \delta} R^{\alpha \beta \gamma
\delta} -
 2 {\hat R}_{\alpha \beta} {\hat R}^{\alpha \beta}  - R_{a b \alpha
\beta} R^{a b \alpha \beta} }
+ 2 {\hat R}_{ab} {\hat R}^{ab}) + {\cal O}(\alpha^{\prime 4})  \Bigr] , \cr}
}
where we
 included the still arbitrary coefficient, $c$,  of the Gauss--Bonnet term
which has  dimensions of (length$)^{-4}$ and which we will fix in a
minute. This expression involves components of the curvature in both the
tangent and the normal bundles. It is strictly-speaking only   valid 
 for totally-geodesic
embeddings of the world-volume (such that $\Omega=0$, see  appendix A) in a Ricci-flat
ambient space-time. We will see below  how 
to  generalize the formula  to other embeddings.

The ambiguous coefficient of the Gauss-Bonnet term can be fixed
by the following
 indirect argument. Consider type IIA  theory compactified on a K3
 surface, which is dual to the heterotic string on $T^4$
 \refs{\hulltown}. A wrapped D4-brane carries a D0-brane charge induced by the
presence of the
gravitational WZ term in the world-volume action \wzdef\ \refs{\bershvaf},
\eqn\zroch{ -\frac{1}{48} \frac{1}{8\pi^2} \int_{K3}  \, {\tr}
 (\R\wedge \R)  = -1 \ .}
Now  comparison with the spectrum of the heterotic string 
leads to  the following mass formula on the type-IIA side \ramwal\
\eqn\kthree{
m = T_{(4)} n_4 V_{K3} + T_{(0)} n_0
}
with $n_4$ and $n_0$ the  D4-brane  and D-particle charge.
 The DBI action contribution to the mass is proportional to the K3
volume
which does not give the required shift. However, this is given by
 including the $R^2$ term in the action.  The presence of this term
 results in a contribution that is proportional to the  Euler class of
 the manifold,
\eqn\euldef{\chi \equiv  \int_{K3}d^4 \zeta \; {\cal L}_{GB}
   =\frac{1}{32\pi^2}\int_{K3} d^4\zeta
{\sqrt { g}}\; {R}_{\alpha \beta \gamma \delta}
{R}^{\alpha \beta \gamma \delta} = 24.}
We use here the fact that K3 is Ricci flat, and that all curvatures
with normal-bundle indices are zero. 
  Substituting in \treelevel\
and using the standard expressions for the tensions $T_{(0)}$ and $T_{(4)}$
gives the correct BPS mass formula for the type IIA theory
compactified on K3, provided we set the coefficient   $c=0$.

We will note two pieces of evidence that give partial
confirmation of our result. First, consider
a  D-particle sitting at a point of a static curved  space-manifold, such as a
Calabi-Yau space. It can be checked that the only potential
 ${\cal O}(\alpha^{\prime 2})$ corrections in this case could come from the contraction of two
  curvatures with all indices in the normal bundle. Such terms are
  explicitly absent in \treelevel, consistently with the fact that the
mass of the D-particle should not receive any $\alpha^{\prime}$  corrections.\foot{We thank
M. Douglas for this argument.} 
The second consistency check   is based on space-time
supersymmetry. Consider a ${\cal N}=4$ 
 supersymmetric compactification of type II theory to four
 dimensions with 16 D3-branes and 64 orientifold three-planes.
 We will
discuss this compactification in more detail in section 3.
 The  gravitational Wess-Zumino term  receives   contributions both from the
D3-branes and from the O3-planes, adding up altogether to
24 times the contribution of one D3-brane \refs{\dasguptaa}.
Supersymmetry  relates these  CP-odd 
Wess-Zumino  terms  to their CP-even counterparts
 (see, for example, \refs{\theisen}). If we assume that the
 coefficients of the CP-odd  $($curvature$)^2$
  terms for an O3-plane and a D3-brane
 are in the same proportion as their  CP-even counterparts, then  the
 coefficient of ${R}_{\alpha \beta \gamma \delta}
{R}^{\alpha \beta \gamma \delta}$ agrees with the one found
above. This confirms independently the choice of the 
Gauss-Bonnet coefficient  $c=0$.

Let us consider finally  the case of  
arbitrary world-volume embeddings,
which are not totally geodesic ($\Omega\not=0$).
We have found that the following lagrangian
\eqn\fullcpeven{\eqalign{
{\cal L}_{CP-even}^{(p)}& = T_{(p)} e^{-\phi} {\sqrt{ g}}
 \Bigl[  1-\frac{1}{24}\frac{(4\pi^2\alpha^\prime)^2}{32 \pi^2}  
\Bigl( ({ R}_T)_{\alpha \beta \gamma \delta} ({ R}_T)^{\alpha \beta \gamma
\delta} - \cr
&   - 2 ({ R}_T)_{\alpha \beta} ({ R}_T)^{\alpha \beta} -
{(R_N)}_{ \alpha \beta a b}{(R_N)}^{ \alpha \beta a b }
+ 2 \bar{R}_{ab}\bar{R}^{ab} \Bigr)  + {\cal O}(\alpha^{\prime 4})
\Bigr] , \cr
}}
generalizes \treelevel\ and 
 reproduces correctly the two other scattering amplitudes, (b) and (c),
computed  in \hashkleb. These are the amplitudes involving open
strings, which correspond to 
geometric brane excitations (but not to world-volume gauge bosons).
In the above lagrangian
 $ R_T$ and $R_N$ are the world-volume curvature and the $SO(9-p)$
gauge field strength, which in the case of a totally geodesic
embedding reduce to the pull-backs of the ambient curvature (see
appendix A), and we have defined 
$\bar{R}_{ab} \equiv {\hat R}_{a b}   +
{g}^{\alpha \alpha^\prime} {g}^{\beta \beta'} \Omega_{a \vert \alpha
\beta}\Omega_{b \vert \alpha^\prime \beta'}$. 
As previously, our  arguments cannot  fix the coefficients of terms
involving the bulk Ricci tensor or the trace of the second fundamental
form  -- both of which vanish  by virtue of the
mass-shell conditions at linearized level. We have not checked whether
 comparison with the above scattering amplitudes
 leaves  any residual ambiguities in the ${\cal O}(\alpha^{\prime 2})$
disk-level  action, beyond those corresponding to 
local redefinitions of the geometric brane coordinates $Y^\mu(\zeta)$.

The CP-odd  $($curvature$)^2$  terms in the disk-level D-brane action
 derived in  \refs{\cheunga}\ must involve closed forms on the
 world-volume, in order to be invariant under gauge transformations of
 the \RR\  potentials. Hence, for non-geodesics embeddings, they must
 also be expressed in terms of the
 world-volume and $SO(9-p)$ curvatures,
\eqn\cpodd{
{\cal L}_{CP-odd}^{(p)} = T_{(p)}\Bigl(  C^{(p+1)} +
{\pi^2\alpha^{\prime 2}\over 24} C^{(p-3)}\wedge \left[ {\rm tr} ({
\R_T}\wedge { \R_T}) - {\rm tr} ({\R_N}\wedge {
\R_N} ) \right]  \Bigr) .
}

We will now focus our attention on non-perturbative corrections to
the D3-brane action. To this end, we will
specialize for simplicity to the case where the ambient space-time
is the direct product of the worldvolume times a  normal space. 
The terms that survive in \fullcpeven\ and \cpodd\ in this special
case  can be combined together as follows

\eqn\simple{\int_{M^{(4)}} {\cal L}_{R^2}^{(3)tree} =
 {1\over 192\pi}
\, \int_{M^{(4)}} \Bigl[ -  \tau_2 \; {\rm tr} (\R\,{\wedge^*\R}) 
+\tau_1 \;  {\rm tr}(\R\wedge \R)\Bigr]  ,}

\noindent where $\tau$
is defined in \scalar\  and $\R$ will stand from now on  for
the SO(1,3)-valued curvature two-form on the world-volume of the D3-brane.

\newsec{Instanton corrections to the D3-brane action}

The conjectured $SL(2,\IZ)$ invariance of type IIB theory requires
that the D3-brane world-volume theory be invariant under $SL(2,\IZ)$
transformations of the supergravity  backgrounds. If one keeps only
the two-derivative terms in the effective action, the invariance is
 guaranteed by  the standard
Montonen--Olive duality of four-dimensional ${\cal N}=4$
supersymmetric Yang--Mills theory.  More generally, it is also true
that the sum
 of DBI and non-gravitational WZ terms of the D3-brane
action exhibits electromagnetic duality
\refs{\gibbons,\tseytlina,\greengutd}. This is consistent with
$SL(2,\IZ)$ invariance in the special limit in which
 gravitational-curvature
 and covariant
`acceleration' terms can be neglected.

The $R^2$ terms obtained from disk amplitudes, on the other hand,
do not by themselves respect  $SL(2,\IZ)$ invariance.
To see why, it is useful to first rescale the metric from string to the
Einstein frame, $G_{\mu \nu}  = e^{\phi/2}G_{\mu \nu}^E$. In this frame
the dilaton drops out of the Nambu-Goto  action, while
the parity-conserving and parity-violating $R^2$ terms are
multiplied, respectively,
by $e^{-\phi}$ and  by  $C^{(0)}$.
Since
 $SL(2,\IZ)$ transformations do not act on the Einstein-frame metric
 whereas  $\tau$  transforms as a modular parameter,
 invariance is explicitly lost. This applies both to the parity
 conserving terms that multiply $\tau_2$ and to the gravitational
WZ terms  that multiply $\tau_1$.

Our purpose in this section is to  explain how  D-instanton corrections to the
effective D3-brane action  restore invariance under $SL(2,\IZ)$.
One way
 to understand the presence of instanton corrections is to compactify a
 transverse direction and then perform a T-duality tranformation that
 changes the D3-brane into a D4-brane. Now the D4-brane forms
 threshold bound states with D-particles \polch, whose Euclidean
 trajectories can wrap around the compact dimension. Taking these
 Euclidean trajectories into account is equivalent in T-dual language
 to summing the D-instanton corrections.
As it turns out  the effective Wilsonian action with its
instanton corrections still leads to an anomalous phase of the
partition function under modular transformations. This phase cancels,
however, precisely  the anomalous contribution  of the massless
${\cal N}=4$ vector multiplet living on the D-brane, provided also that
the metric background satisfies an appropriate consistency condition.

\subsec{Duality  between type IIA on $K3 \times T^2$ and F-theory on
$K3\times T^4$}

 The D-instanton corrections to the $R^2$ terms
of the  D3-brane action are, in the  special case
of a
trivial normal bundle discussed here,
 summarized by  \fullconj\ . \foot{A  non-trivial normal
bundle poses some extra difficulties which we will comment on
briefly later.}
More explicitly, the effective Wilsonian action contains  a
 CP-even piece
\eqn\cpeven{
\int_{M^{(4)}} {\cal L}_{R^2}^{(3)\  CP-even}
  =  {1\over 32 \pi^2} \int_{M^{(4)}}
 \log |\eta(\tau)|^2 \; {\rm tr} (\R\, {\wedge^*\R}) }
where the logarithm of the Dedekind function has the weak-coupling  expansion,
\eqn\etax{
\log  \vert \eta (\tau)\vert^2  = -{\pi \over 6} \tau_2 -
  \left[q + {3 q^2 \over 2} + {4 q^3\over 3} \cdots\ +  {\rm cc}  \right].
}
The first term is the coefficient of the
perturbative disk diagram contribution,
 and the series of powers of $q=e^{2\pi i \tau}$ are the
D-instanton corrections. Likewise, the
 CP-odd piece of the effective Wilsonian action  reads
\eqn\cpodag{\int_{M^{(4)}} {\cal L}_{R^2}^{(3)\  CP-odd}  = -{i\over 32\pi^2}
\int_{M^{(4)}}  \log{\eta(\tau)\over \eta({\bar \tau})}\; \tr(\R\wedge
\R) .}
where the  logarithm may be expanded as
\eqn\etaex{\log{\eta(\tau)\over \eta({\bar \tau})} = {i\pi \over 6}
\tau_1  - \left[q + {3q^2\over 2} + {4q^3\over 3} + \cdots\ - {\rm cc}
\right]. }
The first term, which again arises from the perturbative disk diagram,
 coincides with the $C^{(0)}$ term in \cpodd\ (for $p=3$ and when
the normal bundle is trivial) whereas  the series of powers of $q$
stand  again for the D-instanton corrections.
The derivative with respect to $\tau$ of this expression can be
written in  terms of the imaginary part of the second Eisenstein
form,  which is holomorphic but not modular covariant.

One way of arriving at the above action,  outlined
in reference \dasguptaa, can be rephrased in terms of the duality
between
F-theory compactified on $K3 \times T^4$ and  type-IIA theory
compactified on $K3 \times T^2$, which is also dual to heterotic theory on
$T^6$.  The exact expression for the $R^2$ terms in the type IIA theory can
be derived from a one-loop calculation
\refs{\harvmoor,\gkkopp} and is a function of the
torus  \kh\
modulus $T$.  On the F-theory side, on the other hand,
these terms come from the sum of
the   actions on 24 $(p,q)$ seven-branes compactified to
four dimensions on $T^4$.  A maximum of eighteen of these may be
seven-branes
with the same values of $p$ and $q$, but fortunately  the $R^2$ terms of interest
are the same for all types of seven-branes. Furthermore, 
a  T-duality
 in the four $T^4$ directions  maps these $R^2$
terms for a $(p,q)$ seven-brane to the corresponding terms  for a
D3-brane.  This chain of dualities
therefore predicts  that
the $R^2$ terms in the D3-brane action are exactly $1/24$th of the
corresponding type-IIA loop amplitude, if one identifies $\tau$ in the
former with
the \kh\ modulus $T$ in the latter. Comparing with the exact type-IIA
expression \refs{\harvmoor,\gkkopp} leads then to the conjectured
action  \fullconj\ on  an individual D3-brane.

Let us explain this argument in some more detail. In the case of
a D7-brane, the disk-level action contains the terms
\eqn\dseven{
\int_{M^{(8)}} {\cal L}_{R^2}^{(7)}
  = {1\over\lambda_s^B} {1\over 192\pi}{1\over (4\pi^2\alpha^\prime)^2}
 \int_{M^{(8)}} \; \left(
- e^{-\phi} \; {\rm tr} (\R\, {\wedge^*\R}) + C^{(4)}\wedge {\rm tr} (\R\,
 {\wedge \R}) \right)\ .
}
Switching from the string to the ten-dimensional
Einstein frame eliminates  the dilaton
dependence of this action. Since both the Einstein-frame metric and
the four-form $C^{(4)}$ are inert under $SL(2,\IZ)$ transformations,
we can conclude  that  \dseven\  is $SL(2,\IZ)$ invariant  and must
thus describe  the $R^2$ terms in the world-volume action of any
$(p,q)$  seven-brane. After
compactification on $T^4$ the above seven-brane action
reduces to the leading, disk-level term of the D3-brane action
\fullconj\ provided we identify the
modular parameter
\eqn\parm{
\tau = {(4\pi^2\alpha^\prime)^{-2}\over \lambda_s^B}\;
\left(  \int_{T^4} C^{(4)} + i\; v^{(4)}_E \right)
}
where $v^{(4)}_E$ is the volume of the four-torus in the ten-dimensional
Einstein frame. It can be checked that this identification is the one
dictated by the T-duality transformation that maps the seven-branes to
D3-branes.

>From the seven-brane point of view the instanton corrections arise  from
Euclidean  trajectories of D3-branes wrapped around  the compact $T^4$
 and the quantity $q = e^{2\pi i\tau}$ is
the semi-classical measure corresponding to a singly-wrapped
trajectory.\foot{ See
\refs{\insta,\inst,\instt,\instn,\instb} for
 discussions of  such extended D-brane instantons in other
contexts.}
It is worth noting that in the eight-dimensional
 compactification of F-theory both
the dilaton and the conformal factor of the metric of the $\sigma$-model
that is induced on the seven-branes
vary as functions of the transverse
two-dimensional space. This variation however precisely cancels out
(see for instance \refs{\ab}) when one considers the Einstein
metric, which is consistent with the fact that the modulus $v^{(4)}_E$
should  not depend on the precise locations   of the seven-branes.

It is  important to check that the instanton  corrections to the
Wess Zumino action  do not affect the anomaly inflow argument of
\refs{\ghm}. For example, consider the configuration in which a
D3-brane intersects a D7-brane on a line, along which chiral fermions
propagate.  The  consequent gravitational anomalies in the intersection 
domain  are supposed
to be cancelled by anomalous inflow from the branes.
The anomalous variation of the WZ action of the D3-brane under
space-time reparametrizations  depends only on the source term in
the anomalous Bianchi identity for the  \RR\  one-form field
strength. But this is not affected by the non-perturbative D-instanton
corrections because the semi-classical  measure $q$ has trivial
monodromy around the D7-brane,
 \eqn\bian{ \int_{S^1}\; d \; {\rm
log}{\eta(\tau)\over \eta(\bar\tau)} = {i\pi\over 6\lambda_s^B}\;\int_{S^1}\;
 d \; C^{(0)}\  = {i\pi\over 6}\ ,}
where  $S^1$ is a circle around the D7-brane.
The anomalous variation in the
intersection domain therefore comes entirely from the disk-level
part of the D3-brane WZ action.   A similar argument should ensure
that instanton effects in the D7-brane do not affect the inflow
argument.

As we will explain in the concluding remarks, a  chain of
reasoning analogous to   the one described in this subsection also requires
the presence of nonperturbative  $R^4$ contributions  to the
seven-brane action, coming from localized D-instantons.

\subsec{Modular anomalies}

The elements of $SL(2,\IZ)$ are generated by
the transformations $T$, which acts as a shift  $\tau\to \tau+1$,
and $S$, which acts as the inversion $\tau \to -1/\tau$.  The
corresponding transformations of the logarithm of the Dedekind function are 
\eqn\vareta{\log{\eta(\tau+1)} =
\log{\eta(\tau)} + i{\pi\over 12},
\qquad
\log{\eta(-1/\tau)} = \log{\eta(\tau)} -  i{\pi\over 4} +
\frac{1}{2}\log{\tau}.}
As a result, the effective Wilsonian action \fullconj\ is not by
itself $SL(2,\IZ)$ invariant. This situation is familiar from other
related examples 
 \refs{\dfkz,\carlustov,\kaplouis,\kaplouisp}. The $\tau$-dependent
 variation of the Wilsonian action  should cancel the anomalous
contributions of
 the massless modes, which are  the gauge boson  and its  fermionic
 partners in the abelian ${\cal N}=4$ vector multiplet that describes the
 low-energy dynamics of the D-brane.  The constant part of the
 variation \vareta\ on the other hand, leads to a net anomalous
 variation  equal to the integral of  a topological
 density-- the first Pontryagin class of the  manifold wrapped by the
 world-volume of the D3-brane.
 This is  an integer multiple of $2\pi$ in
 consistent backgrounds, for which the
 first Pontryagin class is an integer multiple of 24.

To understand the modular anomaly of the massless D3-brane modes, we need
to consider the action for ${\cal N}=4$ supergravity coupled to the
${\cal N}=4$ Maxwell theory \refs{\deroo,\bergshoeff}. Both the
Maxwell field and its fermionic partners contribute anomalous
variations of the partition function. 
One way of analyzing the  anomaly of the fermions is to start with a
 gauge-invariant description of the IIB theory in which there are
 three scalar fields, $\tau$, ${\bar \tau}$ and $\phi$,
 that parameterize the group manifold of $SL(2,\IR)$.  These package
 into a zweibein, $V_i^a$, which is an $SL(2,\IR)$ matrix
\eqn\vdef{V_i^a = {1\over \sqrt{\tau_2}} \pmatrix{\tau_2 \cos \phi + \tau_1
\sin\phi & -\tau_2 \sin\phi +\tau_1 \cos\phi \cr
\sin \phi & \cos\phi \cr}.}
The global $SL(2,\IR)$  acts by matrix multiplication on $V$ from the
 left while the local $U(1)$ acts from the right.  The $U(1)$ gauge
 potential is the composite field
\eqn\pot{
a_\mu = \partial_\mu \phi - {\partial_\mu \tau_1\over 2\tau_2}.
}
  The field $\phi$ transforms as \eqn\phitrans{\phi(x) \to
\phi(x) + \Sigma(x),} under the local $U(1)$ (where $0\le \Sigma \le
2\pi$ is the gauge parameter) and by
\eqn\morphi{\phi \to
\phi  -  {i\over 2} \log\left({c\tau+d \over c {\bar \tau} +d}\right)}
under the $SL(2,\IR)$ transformation which acts on the field $\tau$ in the usual
manner,
\eqn\modtrans{  \tau \to A\tau =  {a \tau + b \over c \tau + d} \qquad
\hbox{where} \quad A= \pmatrix{a & b \cr c & d}.}
Notice that $a_\mu $  is invariant under $SL(2,\IR)$.

This  local $U(1)$ also acts as a  phase rotation on the four Weyl
 fermions
 whose  charge is
$-1/2$. On the other hand the  fermions are inert under
$SL(2,\IR)$.
In the absence of anomalies the field $\phi$ can be eliminated by
fixing the  gauge, and then the remaining scalars $\tau$ and
$\bar \tau$ parameterize the coset $SL(2,\IR)/U(1)$ where
the $U(1)$ is identified with a compact $SO(2) \equiv U(1)$ subgroup of
$SL(2,\IR)$.  At first this appears to be impossible, even in the absence of
D-branes,
because of an apparent anomalous coupling of the  $U(1)$ current
to four gravitons and the $U(1)$ field strength.  This arises from an
anomalous hexagon diagram with circulating chiral
gravitini and dilatini.  However, the  anomalous $U(1)$
transformation of the fermion
determinant can be cancelled by a local counterterm proportional to
$\int \phi \; {\rm F} \wedge {\rm tr}\R^4$ \greengaberd.
The modular anomaly in ten dimensions arises because of the
transformation  of $\phi$ under $SL(2,\IZ)$ which leads to an anomalous
 variation of
this counterterm.

A similar phenomenon occurs on the
world-volume of the D3-brane where  a triangle anomaly
arises from the one-loop diagram  coupling the divergence of the
$U(1)$ current to two gravitons. The world-volume gauginos  which
have a chiral coupling  to the $U(1)$ are
circulating inside  the loop.   The anomalous phase of the partition
function is given by the standard expression
\eqn\uoneanom{ \delta  S_{\rm gauginos} = - {4}\times {1\over 2}   
\int_{M^{(4)}} \; \Sigma(x)\;
 I_{1/2}(R) \,   = {1\over 96 \pi^2}\, \int_{M^{(4)}}\; \Sigma(x)\;
 {\rm tr} (\R \wedge \R )\, ,}
where $I_{1/2}(R)$ is the Dirac index and
the overall coefficient follows from  the $U(1)$ charge of
the four  Weyl fermions. This $U(1)$ anomaly can be  cancelled
by the addition of a  counterterm in the world-volume action
\eqn\cancelterm{\int_{M^{(4)}} {\cal L}^{(3)}_{\rm counter} =
 - {1\over 96\pi^2}\, \int_{M^{(4)}}  \phi\; {\rm tr} (\R\wedge
 \R ).}
Since $\phi$ is an angular variable this expression is unambiguous
only if the first Pontryagin class is an integer multiple of 12  --
 a milder  restriction on consistent backgrounds than the one that we
 will encounter shortly.
Although \cancelterm\  cancels the local $U(1)$
anomaly it violates the symmetry of the action under $SL(2,\IZ)$
transformations since $\phi$ transforms as in \morphi.  The $T$
transformation is nonanomalous but under an $S$ transformation
\cancelterm\ transforms by
\eqn\transterm{
\delta  S_{\rm gauginos}  =
{i\over 192\pi^2}\; \int_{M^{(4)}}\; \log\left({\tau\over \bar\tau}\right)
 {\rm tr} (\R\wedge \R )
,}
which is therefore the fermionic contribution to the modular anomaly.
Put differently, we can use the gauge symmetry to eliminate the
auxiliary field ($\phi =0$), but  then an $SL(2,\IZ)$ transformation must
be accompanied by a compensating $U(1)$-gauge transformation, under
which the fermionic determinants would have exactly the same anomalous
variation as above.

This fermionic contribution only cancels part of the $\tau$-dependence
in the modular transformation of \cpeven\  and \cpodag.
An additional contribution to the modular anomaly arises from the
action of $SL(2,\IZ)$ on  the world-volume Maxwell field strength
\refs{\witvafa,\wittverl}. The Maxwell field is inert under the local
$U(1)$, so one must calculate directly the anomaly in  the current of
the  $SL(2,\IR)$ symmetry which is a good global symmetry of the
low-energy  equations of motion. It should be possible to
calculate this anomaly  as  a
triangle diagram with two external gravitons and one insertion of the 
$SL(2,\IR)$  current. Here we will instead use
directly the results of \refs{\witvafa,\wittverl} for the anomalous
transformation of the Maxwell field partition function.
As with the fermion determinant, the $T$
transformation is nonanomalous but the $S$ transformation leads to the
anomalous
phase,
\eqn\uone{\delta  S_{\rm Maxwell} =
\frac{1}{32 \pi^2}{\Re} \int_{M^{(4)}}\; \log \tau \; \Bigl[
-{1\over 2}\; \epsilon_{\alpha\beta\gamma\delta}
\R^{\alpha\beta}\wedge \R^{\gamma\delta}+ \frac{2i}{3} {\rm tr}
(\R\wedge \R )
\Bigr]\ .
}
This combines with the contribution of the fermion determinant,
\transterm,  to cancel the $\tau$-dependent part of the variation of
the CP-odd term \cpodag, which is the contribution of  the massive string
modes. Likewise, the anomalous phase of the Maxwell partition function
cancels the variation of the CP-even term \cpeven\ for Ricci-flat
backgrounds for which ${1\over 2}\; \epsilon_{\alpha\beta\gamma\delta}
\R^{\alpha\beta}\wedge \R^{\gamma\delta} = {\rm tr}(\R\wedge {^*\R})$.
Recall that our D3-brane action was only obtained modulo terms
involving the Ricci tensor of the bulk.

The modular transformation \vareta\ has also a
 $\tau$-independent part which  gives
rise to  an anomalous phase equal to a constant
multiple of $\int {\rm tr} (\R\wedge \R)$.
For consistency this must be  an  integer  multiple of $2\pi$.
 This condition is fulfilled by consistent
compactifications for which the first Pontryagin class is an integer
 multiple of 48,
\eqn\conn{ {1\over 48} { p}_1(R) = -
\frac{1}{48}\frac{1}{8\pi^2}\int {\rm tr}  (\R\wedge \R)
\in \IZ\ . }
This condition is fullfilled  in particular in the special case of a
 K3 surface. 

It is also interesting to understand how the modular anomaly cancels in
the F-theory  background of section 3.1, or equivalently  the heterotic 
theory compactified
on $T^6$. We have already argued that the  net
contribution to the $\tau$-dependent part of the
anomaly from massive string modes is equal to  24
times that of a single D3-brane,
in accordance with the result of \refs{\harvmoor,\gkkopp}.
 This should cancel the $\tau$-dependent part of
the  modular anomaly
coming from the ${\cal N}=4$ supergravity coupled to $U(1)^{22}$
super-Yang-Mills theory, which is the effective theory  at
 a generic point of moduli
space. To see how this works out precisely, one must use the fact that
 the $U(1)$ charges of the gravitini,
dilatini and gaugini are, respectively, ${1\over 2}$, $-{3\over2}$ and
$-{1\over 2}$, and that the dilatini have opposite chirality compared
to the other fermionic fields  \refs{\deroo,\bergshoeff}. 
Furthermore, the expression of the Dirac index
for a spin $3 \over 2$ field is \refs{\alvarezginsparg}
\eqn\dirac{
I_{3/2}(R) = I_{1/2}(R){\Bigl( D-1 - {\rm
tr}{(e^{i\R/2\pi} - {\bf 1})}\Bigr) } = -21\; I_{1/2}(R) + \cdots 
}
where in the second equality we consider only the four-form relevant 
for   $D=4$ space-time dimensions. Finally the chirality of the
graviphotons is opposite  to that of the gauge
bosons. Putting  all this together one finds the following
contributions to the modular  anomaly, in units in which the
contribution of  an entire  ${\cal N}=4$
vector multiplet is one,
\eqn\anomtot{ \underbrace{22\times {2\over 3}}_{\rm gauge\;  bosons} -
\underbrace{22\times  (-{1\over 2})  \times {4\over 6}}_{\rm
gaugini}  -\;
\underbrace{ 6\times {2\over 3}}_{\rm graviphotons}
\;+\; \underbrace{  (-{3\over 2})\times {4 \over 6} }_{\rm dilatini}
\;+\;
\underbrace{21\times {1\over 2}\times {4\over 6} }_{\rm
gravitini} \ .
 }
The total is indeed 24, consistently   with the
 absence of anomalies as well as with the heterotic/type IIA/F-theory
 duality conjectures.

\newsec{M-theory interpretation}

Another strategy for obtaining the nonperturbative $R^2$
contributions   is  to make use of the dualities that relate string
theory to M-theory.  For example, the bulk $R^4$ terms in the
effective action of ten-dimensional type II string theory can be
obtained from  the one-loop Feynman diagrams that contribute to
four-graviton scattering in eleven-dimensional supergravity
compactified on $T^2$ \refs{\ggv}. The quanta circulating around the
loop are interpreted as D-particles in the type IIA and D-instantons
in the type IIB descriptions of the theory.  The  correspondence
requires the identification of the complex IIB coupling $\tau =
\tau_1+ i\tau_2$ with the complex structure of the two-torus in
M-theory, 
\eqn\corr{
\tau \ \longleftrightarrow\  U=   {R_9\over R_{11}} e^{i\theta}\ ,
}
where $\theta$ is the angle of the torus whose cycles have length
$2\pi R_9$ and $2\pi R_{11}$.
This identification follows from the standard relations connecting  
type-IIA theory and M-theory:
\eqn\mtwo{
R_{11} = l_s \lambda_s^A\ , \ \ \ l_P =
 \left( {2\kappa^2_{(11)}\over (2\pi)^8} \right)^{1/9} = \; l_s\; 
(\lambda_s^A)^{1/3}\ , 
}
as well as  the T-duality transformation relating the IIA and IIB descriptions:
\eqn\tdual{
r^B = l_s^2/r^A\ , \ \ \lambda_s^B = \lambda_s^A \sqrt{r^B\over r^A}\ .
}
Here $r^A=R_{9\ } {\rm sin}\theta$ and $r^B$ are the radii of the ninth 
dimension in the IIA and IIB  respectively  descriptions,
$l_s=\sqrt{\alpha^\prime}$ is the string length scale  
 and $l_P$ the
eleven-dimensional Planck length.
 The expectation value of the \RR\  scalar is $C^{(0)}= {\rm
cot}\theta$. The
ten-dimensional type IIB string theory is   recovered in the limit 
of vanishing $T^2$  volume, $V^{(2)} = 4\pi^2 R_9 R_{11} {\rm
sin}\theta \to 0$, with fixed complex structure.

As  we will now show  the $R^2$ term in the CP-even part of the D3-brane and
D4-brane actions can  also be obtained by an extension of this
argument.  We  consider M-theory in the presence of an M5-brane
that is wrapped around the cycles of a two-torus.
 This is interpreted as a D4-brane in the type IIA description
and a D3-brane in the type IIB interpretation.  We will consider a
one-loop diagram with two external gravitons scattering in the
world-volume of the M5-brane.  The circulating quanta with
non-vanishing momentum in the eleventh direction are now
interpreted as bound states of D-particles  with the D4-brane.  In the
type IIB description they correspond to the D-instantons in the
D3-brane whose contributions have been resummed  in expression \fullconj.

This one-loop effect can be in principle computed as  the sum of Feynman
diagrams with the  component fields of the chiral $(2,0)$
tensor multiplet of the M5-brane circulating in the loop.  A more
efficient method for obtaining this expression is to make use of the
manifestly supersymmetric first-quantized light-cone gauge formalism
of \refs{\ggv}\ in which the amplitude is described as a trace over
the states of the superparticle circulating around the loop and coupled to
two external gravitons. This method, inspired by the string-theory
calculation, will be sufficient to find the
CP-even part of the action.  The eleven-dimensional coordinates are
labelled $x^0, \dots x^9, x^{11}$ and  the light-cone gauge is defined
so that  $X^+(t) = (X^0(t) + X^1(t))/\sqrt 2 = p^+ t + x^+$.  The
two-torus will be chosen to be in the $x^9,x^{11}$ directions while
the M5-brane will be oriented in the $x^0, x^1, x^2, x^3, x^9, x^{11}$
directions. The light-cone gauge vertex operator of a graviton with
transverse momentum $p_i$  ($i=2,\dots,9,11$) and
polarization $\epsilon_{ij}$ is \refs{\ggk}
\eqn\vertdef{
V(\epsilon, p)  = {2 \pi} \left({1\over 32} {\cal S}
 \gamma^{il} {\cal S} p_l \;  {\cal S}
 \gamma^{jm} {\cal S} p_m  \;+\; \cdots  \right)\;\epsilon_{ij}\; 
 e^{ip \cdot X},}
(where it is also necessary to set $p^+ =0$ as usual in defining
 light-cone gauge vertex operators). 
The real $SO(9)$ spinor  ${\cal S}^a$ ($a=1,\dots,16$) is the
 $\gamma^+\gamma^-$ projection of a covariant $SO(10,1)$ spinor and it
 satisfies the anticommutation relations  $\{S^a, S^b\} =
 \delta^{ab}$. We have only indicated here the term in $V(\epsilon, p)$
 which is  quartic
in the spinor coordinates, since other terms do not contribute to the
amplitude of interest. We have also 
 fixed the overall
 normalization so as to make contact with the string-theory amplitude
 of section 2.

In evaluating the loop amplitude, we can restrict ourselves to  the special
kinematic setup in which the polarizations of the two external
gravitons, as well as their momenta, are in the directions $x^2, x^3$
 which are the world-volume directions transverse to the
light-cone and $T^2$ directions.  This will be sufficient if we are
not interested in the normal bundle. More generally, it is sufficient
that the polarizations and momenta have no components  in the eleventh
dimension. 
The expression for the loop amplitude is~:
\eqn\ampli{
A(\epsilon_1, p_1; \epsilon_2, p_2)  = \frac{1}{{2 V}^{(2)}}
\sum_{l_9,l_{11}}  \int {d^{4}k\over (2\pi)^4}  \   
\Tr_{{\cal S}}\prod_{r=1,2}
\int_0^\infty  dt_r V(\epsilon_r,p_r)  e^{-\pi (G^{IJ} l_I l_J + k_r^2)\; t_r},
}
where the trace is over the fermionic modes and $r=1,2$ labels
the  two graviton vertices.   The
four-dimensional continuous momenta in the legs 
of the loop are denoted  $k_r = k + \sum_{s=1,r} p_r$  and $G^{IJ}l_Il_J$
is the square of  the Kaluza-Klein momentum  in the compact directions,
$x^9$ and $x^{11}$.   This special configuration of momenta and
polarizations is sufficient to determine the parity-conserving part of
the amplitude.  However, a more general configuration would be
necessary to extract the parity-violating piece which is proportional
to the world-volume form  and has vector
indices associated with all four world-volume directions.

We will first discuss the trace over the fermionic modes, ${\cal S}$.
The  M5-brane breaks half of the thirty-two supersymmetries,
preserving those of definite chirality 
$\gamma^0 \gamma^1 \gamma^2 \gamma^3  \gamma^9
\gamma^{11} $. Decomposing the $SO(9)$ spinor ${\cal S}$ into 
two eight-dimensional Majorana-Weyl $SO(8)$ spinors  $S_R$ and
$S_L$, this relation translates into the usual relation between right
and left moving spinors in the type II theories in the presence of a
D-brane,
\eqn\rels{ S_L = \gamma^2 \gamma^3 \gamma^9  S_R \equiv \gamma^{(3)}
S_R,}
Substituting this decomposition into the trace, the kinematic factor
takes the form
\eqn\strax{
K(1,2) = {1\over 16}\; \Tr_{{S_R}} \prod_{r=1,2}  \epsilon_{r,ij}\;
p_{r,l}\; p_{r,m}\;  (S_R \gamma^{il} S_R)\, (S_R
\gamma^{(3)}\gamma^{jm}\gamma^{(3)} S_R).}
Now one can use the relation  $\gamma^{(3)}\gamma^{j
m}\gamma^{(3)} = \gamma^{{j'} {m'}} D_{j'}^{\ \ j} D_{m'}^{\ \ m}$, 
with $D$  the matrix introduced in section 2, and the well known fact
that  the trace over eight $S_R$'s is proportional to the $t_8$
tensor. The final result 
agrees precisely with  the kinematic factor $K(1,2)$ given in
eqs. (2.6-8).

Next, we shall evaluate  the integrals over the momenta and over the
two proper times
$t_r$.  Since we are only interested in the four-derivative terms, we
can set to zero the external
 momenta in the 
integrand. Then, letting $t\equiv t_1+t_2$,  and integrating over the loop
momentum $k$  and  over the difference $(t_1-t_2)/2$  we find 
\eqn\nolasde{
A(\epsilon_1, p_1; \epsilon_2, p_2) = 
\frac{ K(1,2)}{128\pi^2  V^{(2)}} \int_0^\infty{ \frac{dt}{t}
\sum_{l_9,\, l_{11}} e^{ -\pi t \vert l_9 -
l_{11}\tau \vert^2 / {v^{(2)} \tau_2} }
}\ .
}
We have here used  the explicit form for  the square of the 
 Kaluza-Klein momentum 
\eqn\invdef{G^{IJ}l_I l_J =
\frac{4\pi^2}{{V}^{(2)} \tau_2} {|l_9- l_{11} \tau |}^2,}
 where $\tau = U$ is the
complex structure of the torus, and we have absorbed a factor of
$4\pi^2$ by a  rescaling of $t$. 
Poisson resumming  over $l_9$ leads to 
\eqn\lelevennotzero{
A(\epsilon_1, p_1; \epsilon_2, p_2) =
 {K(1,2)\over 128 \pi^2} {\left( {\frac{\tau_2}{V^{(2)}}}\right)}^{1/2}{  
\int_0^\infty  \frac{dt}{t^{3/2}} \sum_{w_{9},\, l_{11}} e^{
-\pi\tau_2  \left( 
 w_9^2 v^{(2)}/ {t} +  l_{11}^2 t /{v^{(2)}}\right) - 2\pi i\tau_1
l_{11}w_9 }}\ ,
}
which is the  starting point for analyzing the various contributions.

 In the analysis that follows, we will see that the Wilsonian action
  in ten non-compact 
  dimensions corresponds to  the $l_{11}\not= 0$ contributions in the
  above sum, that is 
   to non-zero  momentum along the hidden eleventh
  dimension of string theory. On the other hand, the contributions
  with $l_{11}=0$ and $w_{9} \ne 0$ will reproduce
  the one-loop open-string amplitude,  which is saturated by massless
  modes in the ten-dimensional IIA ($R_9\to \infty$)  or  IIB ($R_9\to 0$)  
  theory. We may further separate the $l_{11}\neq 0$ sum into
  contributions with $w_9\neq 0$ and $w_9 = 0$. The terms with
  $w_9\neq 0$ give a 
\eqn\landnnotzero{\eqalign{
{\cal A}^{\rm non\; pert}= {1\over 128\pi^2}
{\left({\frac{\tau_2}{V^{(2)}}}\right)}^{1/2} & \int_0^\infty{ \frac{dt}
{t^{3/2}} \sum_{w_9 \neq 0,\, l_{11} \neq 0} e^{ -\pi \tau_2 \left( 
v^{(2)} w_9^2 / t +  l_{11}^2 t /{v^{(2)}}\right) - 2\pi i
l_{11}w_9 \tau_1}} \cr
&=  \frac{\tau_2^{1/2}}{64\pi^2 V^{(2)}}  \sum_{w_9 \neq 0,\,
l_{11} \neq 
0}  {\left|{\frac{l_{11}}{w_9}}\right|}^{1/2}
K_{1/2}(2\pi |w_9 l_{11}| \tau_2)\;  e^{-2\pi i w_9 l_{11} \tau_1} \cr
&= \frac{1}{64\pi^2 V^{(2)}}  \sum_{N=1}^\infty \; \sum_{n\vert N} 
{1\over n}\; \left( e^{2\pi i N\tau} + {\rm c.c.}\right)
}}
\noindent 
This reproduces precisely the non-perturbative D-instanton sum in the
expansion \etax\ of the CP-even part of the effective action,
and gives an independent confirmation of our non-perturbative conjecture.

   The terms with $w_9=0$ and any value of $l_{11}$ (including the
 term with $l_{11}=0$) are quadratically ultraviolet divergent.  The
 total divergence may be isolated by converting the sum over the
 Kaluza--Klein charge, $l_{11}$, to a sum over winding number in the
 eleventh dimension, $w_{11}$, by the Poisson  resummation, 
\eqn\wzero{\eqalign{{\cal A}^{\rm tree}= {1\over 128\pi^2 }
{\left({\frac{\tau_2}{V^{(2)}}}\right)}^{1/2} \int_0^\infty{ \frac{dt}
{t^{3/2}} \sum_{l_{11}} e^{ -\pi \tau_2 l_{11}^2 t/v^{(2)}}
}&  = {1\over 128\pi^2} \int_0^\infty{ \frac{dt}
{t^{2}} \sum_{w_{11}} e^{ -\pi v^{(2)} w_{11}^2/\tau_2 t}} = \cr = &  
\frac{1}{64\pi^3} \zeta(2) \frac{\tau_2}{V^{(2)}} + 
{1\over 128\pi^2}  \int_0^\infty {dt\over
t^2}\cr
}}
The finite piece reproduces the tree-level part of
the effective D-brane action (noting 
$\zeta(2) = \pi^2/6$ and that the M-theory amplitude must be multiplied
by the size of the hidden dimension, $2\pi R_{11}$ in order to make the
comparison with the string amplitude).

The  divergence is now isolated entirely in the zero winding number
term, $w_9=w_{11}=0$.  In any   
microscopic definition of M-theory this divergence would be regularized
and replaced by a finite value.  However, we can argue that its 
regularized value must be zero by an elementary use of duality.   
This follows from the fact that in the decompactified theory ($R_9, R_{11} 
\to \infty$) the  divergent term corresponds to a  $R^2$ term in the effective
five-brane action, proportional to $1/l_p^2$ which is the only scale
in the theory. Compactifying one large dimension ($x^{11}$) would then
give a $R^2$ term  proportional to
\eqn\dimen{
R_{11}/l_p^2 =  (\lambda_s^A)^{2/3}/l_s\ .
}
Since a fractional power of the string coupling constant
  has no sensible interpretation in  the IIA theory, we
conclude that the divergent piece in \wzero\ must be discarded.

    The remaining terms in the double sum  \nolasde\ that have not
been accounted for are the terms
with $l_{11}=0$ and $w_9\not= 0$. 
These add  up to   
\eqn\irdivergent{\eqalign{ {\cal A}^{\rm one\; loop}= {1\over 128\pi^2}
{\left({\frac{\tau_2}{V^{(2)}}}\right)}^{1/2} \int_0^\infty{ \frac{dt}
{t^{3/2}} \sum_{w_9 \neq 0}  e^{ -\pi v^{(2)}\tau_2 w_9^2/  t}}\ ,
}}
corresponding  precisely to the one-loop contributions of the
BPS  open-string modes living on the D-brane. These BPS states consist
of the
massless vector multiplet on the D3-brane, as well as of its 
winding-mode descendants when some  transverse  dimensions are
compact. 
Notice that the $w_9=0$ term in  the annulus diagram 
must be subtracted  in computing  the 1PI  action,  
since it corresponds to a (potential) tree-exchange of a 
supergraviton \refs{\kirb}. 
The sum  \irdivergent\ has a logarithmic divergence which is
responsible for the CP-even  one-loop  modular anomaly  of section 3.

The weakly-coupled   type IIA theory is obtained in the limit $R_{11}
\rightarrow 0$, $R_{9} \rightarrow \infty$, in which the M5-brane
reduces to the D4-brane.  In this case, the instanton terms vanish and
the amplitude reduces to the perturbative terms. 
The weakly-coupled type IIB theory on the other hand is obtained in
the limit of vanishing volume for the two-torus, $V^{(2)}\to 0$ with
$\tau= U$ held fixed. The $N$th term in the non-perturbative sum
\landnnotzero\ comes from the sector of N D-instantons (and 
anti D-instantons)  confined to the world-volume of the brane.
The measure $\sum_{ n|N} 1/| n|$  can be identified with
 the bulk term in the Witten
index  of the T-dual quantum mechanics problem of $N$
D-particles propagating in the world-volume of the D4-brane.  This is
analogous to the 
case of the ten-dimensional D-instanton
 where the measure is $\sum_{n\vert N}1/ n^2$ \refs{\greenguta}.
The extra factor of $1/| n|$
comes from the fact that there are an unequal
number of bosonic and fermionic moduli in the ten-dimensional case
whereas these numbers  are equal in the D4-brane.  The arguments of
\refs{\yi,\sethi,\greenguta} suggest that the
boundary contribution to the Witten index is $1 -\sum_{
n|N }1/| n|$ so that the total Witten index is equal to one.
This agrees with the expectation that there should be threshold bound
states of this system.

In principle, we should also be able reproduce the parity violating
 $\R\wedge \R$ term  in \fullconj\ by considering a loop correction in
 the M5-brane,  but we have not done this. Nevertheless, from
 \fullconj\ and using  the usual T-duality relations between
 the IIB and IIA theories it also follows that the instanton terms
 disappear in  the D4-brane.  In that case, the gravitational contribution to
 the WZ  term in the D4-brane action is simply proportional to
 $C^{(1)}\wedge  {\rm tr}(\R \wedge \R)$, as in \refs{\bershvaf,\ghm,\cheunga}.

\newsec{Comments and discussion}

We have given various arguments for establishing the  $R^2$ terms
  that are relevant to the gravitational interactions of
p-branes in string and M-theory.  These terms are the first nontrivial
corrections in the low energy expansion of  the world-volume action
 and are relevant for  a p-brane in a curved background.
Even at this low order we have left several questions unanswered.
For example, we determined the complete action for geodesic and non
  geodesic world-volume embeddings at tree level (a disk world-sheet),
  including contributions from nontrivial normal bundle.  However, our
  discussion of the nonperturbative effects was restricted to
  situations in which the normal bundle is trivial. 

The discussion  of nonperturbative effects in section 4 was incomplete
 in two respects.  Firstly, we again limited our discussion to
 situations in which the normal bundle is trivial. Secondly, it was
 limited  to the parity-conserving part of the action. These
 restrictions arose, in part, from our use of a light-cone formalism
 which cannot be applied in any obvious manner to the most general
 kinematic configurations. 
Since the M5-brane plays a central and subtle role in the circle of
 dualities it would be interesting to understand in detail how the
 $\R\wedge \R$ terms in the D3-brane can be deduced from a loop
 calculation in the M5-brane. The restriction to configurations with 
trivial normal bundle
   was more than a technical convenience.
  There is a  well-known problem \refs{\witnorm,\bonora,\ruben}
with chiral anomalies that arises when
  embedding the M5-brane in M theory.  After toroidal compactification
  and T duality to the type IIB theory the normal bundle is associated
  with the chiral $SO(6) \sim SU(4)$ R symmetry of the D3-brane embedded
  in ten dimensions.  The possible mixed anomalies in this symmetry
and in $SL(2,\IZ)$
  will require special restrictions on terms in the world-volume action
  that involve the  normal bundle
  curvature. These  should be analogous to the terms involving the
  tangent bundle curvature that we found are necessary for the
  absence of modular anomalies.

Another issue that we have not addressed in any detail is the structure of
higher curvature terms in seven-branes.  In section 3 we saw how the
 $R^2$ terms on a compactified seven-brane are related by T-duality to those of the
D3-brane.  However, we know that there are also $R^4$ terms
 in the seven-brane world-volume action.
According to \wzdef\ the disk-level CP-odd WZ term in
the world-volume action of a D7-brane is of the form $C^{(0)}
\, Y_8(R)$, where $Y_8(R)$ denotes an eight-form that is the linear
combination of  ${\rm tr} (\R\wedge \R\wedge \R\wedge \R)$
and  $({\rm tr} (\R\wedge \R))^2$ that follows from \sqdir.
Similarly, considerations of scattering on the seven-brane
world-volume will lead to CP-even $R^4$ terms. Arguments similar to
those given in this paper imply the presence of nonperturbative contributions to
these $R^4$ terms. The necessity for such contributions can be seen in
particular by considering the F-theory background consisting of 24
$(p,q)$ seven-branes, which is dual to the heterotic or type-I string
compactified to eight dimensions on $T^2$. The known structure of the
$R^4$, as well as the  $R^2 F^2$ and $F^4$ terms in this background 
\refs{\inst,\instn,\instt,\lerche,\kristin,\lerch,\lerc}
implies that in the type-I language the quartic action of
 both orientifold seven-planes and D7-branes
should receive non-trivial  D-instanton corrections.
  Although we will not discuss the details here, it should again be possible
 to motivate the form
 of these  nonperturbative $R^4$ terms by the absence of modular
 anomalies.  Furthermore, as with the $R^2$ terms in the D3-brane,
 the M-theory origin of such terms should emerge from
duality relations between type-${\rm I}^\prime$ 
string theory on $T^2$ or F-theory on K3  and eleven-dimensional
M-theory compactified on $T^2\times I$ with  $I$  the  finite-length interval of
the Horava--Witten
configuration \refs{\horavwit}.  The D-instanton
contributions on the string theory side 
should presumably arise from  a one-loop calculation
of four graviton scattering in the above compactification of M-theory.

Finally, we have not addressed interesting issues concerning the
compatibility of the curvature terms in the D3-brane world-volume
action and the correspondence between type IIB superstrings in
$AdS_5\times S^5$ and ${\cal N}=4$ supersymmetric Yang--Mills
theory.   One obvious  issue is the consistency of the
higher-order curvature terms in the world-volume action of a test
D3-brane probe parallel to the $AdS_5$ boundary. These curvature terms are
a priori nonzero since the  Riemann curvature is nonzero in this
background. However, there are surely other world-volume terms
that we have not addressed, which include powers of the self-dual
fifth-rank field strength, $F_5 = dC^{(4)}$, which is also nonzero
in this background. It would be interesting to check whether 
 these corrections are  consistent with the statement that
supersymmetry protects the energy
per unit volume 
of a stack of N parallel D3-branes.  
Another interesting question is how the D-instanton terms in the
D3-brane action contributions are  interpreted  in terms of
the contributions of Yang--Mills instantons.

\ 

\noindent {\bf Aknowledgements}

We are grateful to E. Cremmer, M. Douglas, K. F{\"o}rger, 
M. Gaberdiel, G. Gibbons, R. Minasian, M. Petropoulos, B. Pioline and 
S. Silva for useful discussions, and to A. Tseytlin for a critical
reading of the manuscript.

\vfill
\eject

\appendix{A}{The geometry of submanifolds}

 We will here collect some standard facts on  the
geometry  of submanifolds which can be found in standard texts, such as
 \refs{\eisen,\koba}.
   The embedding of the D-brane worlvolume in the ambient
spacetime is described by the
   ten  coordinate functions $Y^\mu (\zeta^\alpha)$, with
 $\mu = 0,\cdots, 9$ a space-time index and
   $\alpha=0,\cdots p$ a world-volume index.
 The tangent vectors $\partial_\alpha Y^\mu $
   define a local frame for  the tangent bundle. We may  also define an
   orthonormal frame for the normal bundle,  $\xi^\mu _{a}$
   with $a=p+1,\cdots  9$,  such that
\eqn\norm{
\xi^\mu _{a}\;\xi^\nu_{b}\; G_{\mu \nu} = \delta_{ab}\
\ \ \ {\rm and} \ \ \
\xi^\mu _{a}\; \partial_\alpha Y^\nu\;  G_{\mu \nu} = 0 \ .
}
Both  the $\partial_\alpha Y^\mu $ and the $\xi^\mu _{a}$
transform as  vectors under
target-space reparametrizations. Furthermore the former are vectors of
world-volume reparametrizations, while the latter transform as vectors
under local rotations of the normal bundle.

With the help of these  local frames
 we  can pull back any tensor from the target space-time
onto the tangent and/or normal bundles by contraction. The only
non-trivial pull-back of the ambient metric  is the  induced  world-volume
metric
\eqn\metric{
{ g}_{\alpha\beta} = G_{\mu \nu}\;
\partial_\alpha Y^\mu \partial_\beta Y^\nu
\ .
}
The Riemann tensor, on the other hand, can be pulled back in many
different ways. Using its  symmetry properties,  and the fact that the
sum over a cyclic permutation of three indices is zero, we can write
down the following six independent pull-backs,
\eqn\curv{
R_{\alpha\beta\gamma\delta}\ , \ \
R_{\alpha\beta\gamma a}\ , \ \
R_{\alpha\beta a b}\ , \ \
R_{\alpha \{ ab\} \beta}\ , \ \
R_{\alpha abc}\   \  \ {\rm and} \ \ \ \ \ \ R_{ abcd}\ ,
}
where  the curly  brackets  denote the symmetric combination
of the two normal (and hence also the two tangent) indices.
 These pulled-back Riemann curvatures transform as tensors of
world-volume reparametrizations and normal-frame rotations, but are
scalars under reparametrizations of the target space. We may furthermore
 contract tangent indices with ${ g}_{\alpha\beta}$ and
normal indices with $\delta_{ab}$. Since on the D-brane world-volume
the ambient  metric can be
decomposed as
\eqn\comp{
G^{\mu \nu} = \partial_\alpha Y^\mu 
 \partial_\beta Y^\nu\; { g}^{\alpha\beta}
+ \xi^\mu _a \xi^\nu_b\; \delta^{ab}\ ,
}
 the contraction of spacetime indices before being pulled back
 does not give rise to any new tensors.

  Covariant derivatives can be defined with the  target-space connection
 $\Gamma^\mu _{\nu\rho}$, the
affine world-volume connection  $({\Gamma}_T)^\alpha_{\beta\gamma}$
 constructed out of  the induced metric,
  and the  composite SO(9-p) gauge field
\eqn\gauge{
\omega^{ab}_{ \alpha} = \xi^{\mu , [a}\; \left( G_{\mu \nu}  \partial_\alpha
+  G_{\mu \sigma} \Gamma^{\sigma}_{ \nu\rho}
 \partial_\alpha Y^\rho \right)\;  \xi^{\nu,  b]}\ ,
}
which is defined implicitly by requiring
 that the normal frame be  covariantly constant.
 One important tensor  is  the covariant
 derivative of the tangent frame, also known as the
 `second fundamental form'
 \eqn\second{
 \Omega_{\alpha\beta}^\mu = \Omega_{\beta\alpha}^\mu =
  \partial_\alpha \partial_\beta Y^\mu  -
  ({\Gamma}_T)^\gamma_{\alpha\beta}\;
  \partial_\gamma Y^\mu + \Gamma^\mu _{\nu\rho}\;
  \partial_\alpha  Y^\nu \partial_\beta Y^\rho \ .
  }
It  is a symmetric world-volume tensor and a  vector
of the ambient space-time. Using the fact that ${
 g}_{\alpha\beta}$ is covariantly-constant one can show that the
 tangent-space projection of $\Omega$ vanishes,
$\Omega^\gamma_{\alpha\beta}=0$.
With no loss of  information we may thus  project it on the normal bundle
\eqn\sec{
\Omega_{ \alpha\beta}^a  \equiv  \Omega_{\alpha\beta}^\mu \; \xi^{\nu,
 a}
G_{\mu \nu}\ .
 }
In the special case of a point-particle trajectory,
$\Omega^\mu _{\tau\tau}$
is the covariant acceleration which is everywhere normal
to the world line. A submanifold with zero second-fundamental form is
called auto-parallel or totally geodesic.

The central equations in the theory of submanifolds are the
Gauss-Codazzi equations. They relate the world-volume curvature $
R_T$,  constructed out of the affine connection $\Gamma_T$, and the field
strength
$R_N$ of the $SO(9-p)$ gauge connection $\omega$,  to
 pull-backs of the space-time  Riemann tensor plus combinations of the
second fundamental form,
\eqn\gaus{
({ R}_T)_{\alpha\beta\gamma\delta} =  {R}_{\alpha\beta\gamma\delta} +
\delta_{ab}\left( \Omega^a_{\alpha\gamma}\Omega^b_{\beta\delta}-
\Omega^a_{\alpha\delta}\Omega^b_{\beta\gamma} \right)
}
and
\eqn\gausses{
(R_N)^{\ \ \ ab}_{\alpha\beta} = -R^{ab}_{\ \ \alpha\beta} +  {
g}^{\gamma\delta}\left(
  \Omega^a_{\alpha\gamma}\Omega^b_{\beta\delta}
- \Omega^b_{\alpha\gamma} \Omega^a_{\beta\delta} \right)
\ .
}
Notice that if the ambient space-time is flat these
 world-volume curvatures can be
expressed entirely in terms of the second fundamental form
 $\Omega$. Conversely, if the world-volume is  a totally geodesic
manifold  the
curvature forms of the induced connections coincide with the pull-backs
of the ambient curvature.

\vskip 1cm


\appendix{B} {The systematics of $($curvature$)^2$ terms}

We will here discuss the systematics of possible
 ${\cal O}(\alpha^{\prime 2})$ terms
which can appear as part of the D-brane actions. 
 Invariance under
reparametrizations of the  world-volume and of the ambient space, as
well as under local rotations of the normal frame, 
dictate  that possible  terms are formed by   full contractions of covariant
tensors. At linearized level around flat space and a static D-brane
background the curvature tensor is proportional to a closed-string (graviton)
excitation, and the second fundamental form is proportional to an
open-string (geometric brane) excitation. We assume that all vertices
that do not contain at least two gravitons, four open strings, or one
graviton and two open strings, are protected by supersymmetry and
receive no higher-order corrections. Thus the terms that will interest
us are of the type $R^2$, $R\Omega^2$ and $\Omega^4$ -- all of which
are ${\cal O}(\alpha^{\prime 2})$.

 As discussed in the
main text, our arguments cannot determine any
 terms that vanish by virtue of the
lowest-order equations of motion. These equations impose the vanishing
of the bulk Ricci tensor and of the trace of the second fundamental
form,
\eqn\loweqs{
R^\mu _{\ \nu\mu \rho}= 0 \   \ \ {\rm and} \ \ \ \ \ \Omega^\mu\equiv
g^{\alpha\beta}\Omega^\mu _{\alpha\beta}=0\ .
}
Since  both $R_{\mu\nu}$ and $\Omega^{\mu}$ will
vanish at linearized level by virtue of the mass-shell conditions,
such terms do  not contribute to the  amplitudes of interest.

We will first
consider the case $\Omega^\mu_{\alpha\beta} =0$, 
corresponding to totally-geodesic
embeddings. Since $\Omega$ contains an
open-string excitation at the linearized level, this 
 restriction is
appropriate when  comparing with the two-graviton amplitude (2.5).
One  set of allowed terms are the squares
of the six  curvature  pull-backs \curv,
\eqn\curvv{\eqalign{
&R_{\alpha\beta\gamma\delta} R^{\alpha\beta\gamma\delta}, \quad
R_{\alpha\beta\gamma a}R^{\alpha\beta\gamma a}, \quad
R_{\alpha\beta a b} R^{\alpha\beta a b}, \cr
&R_{\alpha \{ ab\} \beta} R^{\alpha \{ ab\} \beta}, \quad
R_{\alpha abc} R^{\alpha abc}, \quad  R_{ abcd} R^{ abcd}.
}}
\noindent
A second  set of invariants can be constructed by squaring the (partially
or fully)  contracted pulled-back curvatures. 
The partial contractions of the six Riemann curvatures
in \curv\ give rise to six different two-index  tensors.
However, the vanishing of the bulk Ricci tensor imposes three
independent relations among them. We will choose as the three
independent two-index tensors those obtained by contracting tangent
indices, and will denote them  by ${\hat
R}_{\alpha\beta}$, ${\hat R}_{a\alpha}$, and ${\hat
R}_{ab}$ (the hat is there to remind us that
 these tensors are not  the pull-backs of
the bulk Ricci tensor). 
Further contraction of the remaining indices gives rise to a single
independent scalar $\hat R$. In total there are therefore four new
possible  quadratic terms, 
\eqn\curvr{{\hat R}_{\alpha\beta} {\hat R}^{\alpha\beta}, \qquad 
 {\hat R}_{\alpha a}{\hat R}^{\alpha a}, \qquad
{\hat R}_{a b}{\hat R}^{a b}, \qquad  {\hat R}^2.}
Comparison with the two-graviton amplitude fixes the coefficients of
all the terms in  \curvv\   and \curvr,   modulo  one residual ambiguity
corresponding to the Gauss-Bonnet combination.

Let us  turn now to  invariants constructed with  the second fundamental
form $\Omega$.  Recalling that $\Omega$ is traceless,
we find  four  terms of the type
$\Omega^4$, 
\eqn\omegafour{\eqalign{
&(\Omega_{\alpha\beta}\!\cdot \! \Omega^{\alpha\beta})
(\Omega_{\gamma\delta} \!\cdot \! \Omega^{\gamma\delta}), \qquad
(\Omega_{\alpha\gamma} \!\cdot \! \Omega^{\alpha\delta})
(\Omega^{\beta\gamma} \!\cdot \! \Omega_{\beta\delta}), \cr
&(\Omega_{\alpha\beta} \!\cdot \! \Omega_{\gamma\delta})
(\Omega^{\alpha\beta} \!\cdot \! \Omega^{\gamma\delta}), \qquad
(\Omega_{\alpha\beta} \!\cdot \! \Omega_{\gamma\delta})
(\Omega^{\alpha\gamma} \!\cdot \! \Omega^{\beta\delta})
,}}
where we have suppressed in an obvious notation the normal index
of $\Omega$. Likewise, there
are six  invariants of the type  $\Omega^2 R$, 
\eqn\romega{\eqalign{
&R^{\alpha\beta\gamma\delta}
(\Omega_{\alpha\gamma} \!\cdot \! \Omega_{\beta\delta}), \qquad
R^{\alpha\beta}_{\ \  a b}\; \Omega^a_{\alpha\gamma} \Omega^{b\
\gamma}_{\beta}, \qquad 
R^{\alpha\ \  \ \ \beta}_{\  \{a b\}}\;
 \Omega^a_{\alpha\gamma} \Omega^{b\ \gamma}_{\beta}, \cr
&{\hat R}^{\alpha\beta} (\Omega_{\alpha\gamma} \!\cdot \!
\Omega_{\beta}^{\ \gamma}),\qquad\quad
{\hat R}_{a b}\; \Omega^a_{\alpha\beta} \Omega^{b\ \alpha\beta}, \qquad\quad
{\hat R}\; (\Omega_{\alpha\beta} \!\cdot \! \Omega^{\alpha\beta})
.}}
One can use the  Gauss-Codazzi equations to rewrite  some of the above
invariants in terms of the 
 world-volume curvatures $R_T$ and $R_N$. There are four 
possible contractions of these curvatures,
\eqn\curvwo{
(R_T)_{\alpha\beta\gamma\delta}(R_T)^{\alpha\beta\gamma\delta} \ ,\ 
(R_T)_{\alpha\beta}(R_T)^{\alpha\beta}\ , \ 
(R_T)^2\ \ {\rm and} \ \
(R_N)_{\alpha\beta}^{\ \ ab}(R_N)^{\alpha\beta}_{\ \ ab}\ .
}
Four additional invariants are the cross terms involving 
one  world-volume curvature and  the  corresponding pull-back tensor. 
All these terms are linear combinations of the invariants \omegafour\ 
and 
\romega. As discussed in the main text we have found an action that
reproduces correctly the  three scattering amplitudes of section 2. We
did not however analyze systematically the most general combination of the
above  invariants for $\Omega\not= 0$.

  We  have also 
 excluded  from our considerations   terms of
the type $D^2R$, $\Omega D^2\Omega$, $\Omega^2D\Omega$ or $\Omega DR$,
 other than those  obtained after an integration by parts
from  one of the invariants  discussed before.
Such terms, if present,  would have given
${\cal O}(\alpha^{\prime2})$ corrections to the one-point function of
the graviton, the propagator of 
an open string,  the  vertex of three open strings, or to the
mixing of an open with a closed string --  all  of which  are
presumably protected by supersymmetry for some field choice. 
As a check of this  argument we
have verified explicitly that 
$D^2 R$ and $\Omega DR$ terms would, if present, make  unacceptable
 contributions to the three string amplitudes discussed in section
 2.


\listrefs

\end